\documentclass[journal]{IEEEtran}
\IEEEoverridecommandlockouts

\usepackage[normalem]{ulem}
\usepackage[utf8]{inputenc} %

\usepackage[T1]{fontenc}%
\usepackage{amsfonts}
\usepackage{url}
\usepackage{graphicx} %
\usepackage{hyperref} %
\usepackage{dirtytalk}
\usepackage{enumitem}

\makeatletter\let\MYcaption\@makecaption\makeatother
\usepackage[font=footnotesize]{subcaption}
\makeatletter\let\@makecaption\MYcaption\makeatother

\usepackage{textcomp}
\usepackage[style=american]{csquotes}
\usepackage[backend=biber,style=ieee,doi=false,isbn=false,url=false,natbib=true]{biblatex}
\usepackage{booktabs} %
\usepackage{tabularx}
\usepackage{multirow}
\usepackage{afterpage}

\graphicspath{{figures/}}

\usepackage[]{placeins}

\usepackage{nicefrac}       %
\usepackage{microtype}      %
\usepackage{epstopdf}
\usepackage{siunitx}
\usepackage[acronym]{glossaries}

\usepackage{glossaries-extra}

\setabbreviationstyle[acronym]{long-short}

\glssetcategoryattribute{acronym}{nohyperfirst}{true}

\newcommand{\acronymslinktofirstinstance}{%
\renewcommand*{\glsdonohyperlink}[2]{%
    {\glsxtrprotectlinks
        \edef\fieldvalue{\glsxtrusefield{\glslabel}{hastarget}}%
        \ifdefstring\fieldvalue{true}
        {%
            ##2%
        }%
        {%
            \gGlsXtrSetField{\glslabel}{hastarget}{true}\glsdohypertarget{##1}{##2}%
        }%
    }%
}%
}

\newacronym{svd}{SVD}{singular value decomposition}
\newacronym{pca}{PCA}{principal component analysis}

\newacronym{sgd}{SGD}{stochastic gradient decent}

\newacronym{ml}{ML}{machine learning}
\newacronym{fl}{FL}{federated learning}
\newacronym{fedavg}{FedAvg}{federated averaging}

\newacronym{ps}{PS}{parameter server}

\newacronym{ls}{LS}{least square}
\newacronym{lse}{LSE}{least square error}

\newacronym{se}{SE}{square error}
\newacronym{mse}{MSE}{mean square error}
\newacronym{mmse}{MMSE}{minimum mean square error}

\usepackage[british]{babel}
\usepackage{amssymb,amsmath,amsfonts}
\usepackage{amsthm,thmtools}
\usepackage{bm,bbm}
\usepackage{mathrsfs}
\usepackage{mathtools}
\usepackage[ruled, vlined, linesnumbered]{algorithm2e}
\let\oldnl\nl%
\newcommand{\nonl}{\renewcommand{\nl}{\let\nl\oldnl}}%

\SetCommentSty{mycommfont}

\newcommand{\mathpreamblepostcleveref}{
    \crefname{assumption}{Assumption}{Assumptions}
    \Crefname{assumption}{Assumption}{Assumptions}
}

\usepackage{suffix}

\usepackage{array}
\PassOptionsToPackage{table}{xcolor}
\PassOptionsToPackage{dvipsnames}{xcolor}

\usepackage[framemethod=TikZ]{mdframed}

\newcommand\loadpreambletheoremsI{
    \newtheorem{theorem}{Theorem}[section]
    \newtheorem{corollary}{Corollary}[theorem]
    \newtheorem{lemma}[theorem]{Lemma}
    \newtheorem*{theorem*}{Theorem}

    \newtheorem*{corollary*}{Corollary}
    \newtheorem*{lemma*}{Lemma}

    \newtheorem{definition}{Definition}[section]
    \newtheorem*{definition*}{Definition}

    \newtheorem{proposition}{Proposition}[section]
    \newtheorem*{proposition*}{Proposition}

    \newtheorem{assumption}{Assumption}[section]
    \newtheorem*{assumption*}{Assumption}

    \newtheorem{remark}{Remark}[section]
    \newtheorem*{remark*}{Remark}
    \newtheorem{example}{Example}[section]
    \newtheorem*{example*}{Example}
}

\declaretheoremstyle[
	headfont=\bfseries,
	notefont=\mdseries, notebraces={(}{)},
	bodyfont=\normalfont,
	headpunct={\textnormal{:}},
	postheadspace=\newline,
	spacebelow=\parsep,
	spaceabove=\parsep,
	mdframed={
		backgroundcolor=red!02,
		linecolor=red!80,
		linewidth=0.3mm,
		innertopmargin=6pt,
		innerbottommargin=6pt,
		roundcorner=10pt, }
]{red style name}

\declaretheoremstyle[
	headfont=\bfseries,
	notefont=\mdseries, notebraces={(}{)},
	bodyfont=\normalfont,
	headpunct={\textnormal{:}},
	postheadspace=\newline,
	spacebelow=\parsep,
	spaceabove=\parsep,
	mdframed={
		backgroundcolor=blue!02,
		linecolor=blue!80,
		linewidth=0.3mm,
		innertopmargin=6pt,
		innerbottommargin=6pt,
		roundcorner=10pt, }
]{blue style name}

\declaretheoremstyle[shaded={rulecolor=Lavender,
	rulewidth=2pt, bgcolor={rgb}{1,1,1}},spaceabove=6pt, spacebelow=6pt,
headfont=\normalfont\bfseries,
notefont=\mdseries, notebraces={(}{)},
bodyfont=\normalfont,
postheadspace=1em,
]{aaaaa}

\usepackage{etoolbox}
\newcommand{\subscriptifnonempty}[1]{\ifstrempty{#1}{}{_{#1}}}

\DeclareFontFamily{U}{mathx}{}
\DeclareFontShape{U}{mathx}{m}{n}{<-> mathx10}{}
\DeclareSymbolFont{mathx}{U}{mathx}{m}{n}
\DeclareMathAccent{\widecheck}{0}{mathx}{"71}

\makeatletter
\newcommand{\@somedarnpadding}{\mathchoice
  {}%
  {}%
  {}%
  {}%
}
\newcommand{\@somedarnpaddingstar}{\mathchoice
  {}%
  {}%
  {}%
  {}%
}
\newcommand{\@somedarnpaddingwithsubscript}{\mathchoice
  {}%
  {}%
  {}%
  {}%
}
\newcommand{\@somedarnpaddingwithsubscriptstar}{\mathchoice
  {}%
  {}%
  {}%
  {}%
}

\newcommand{\@somedarnafterpadding}{\mathchoice
    {}%
    {}%
    {}%
    {}%
    }
\newcommand{\@somedarnafterpaddingstar}{\mathchoice
    {}%
    {}%
    {}%
    {}%
}
\makeatother

\makeatletter
\newcommand{\cbigoplus}{\DOTSB\cbigoplus@\slimits@}
\newcommand{\cbigoplus@}{\mathop{\widehat{\bigoplus}}}
\makeatother

\makeatletter
\newcommand{\cbigotimes}{\DOTSB\cbigotimes@\slimits@}
\newcommand{\cbigotimes@}{\mathop{\widehat{\bigotimes}}}
\makeatother

\NewDocumentCommand\SetSymbol{o}{%
    \nonscript\:#1\vert
    \allowbreak
    \nonscript\:
    \mathopen{}
}
\DeclarePairedDelimiterX\gullwingsparenthesis[1]{\{}{\}}
{%
    #1%
}
\DeclarePairedDelimiterX\squareparenthesis[1]{[}{]}
{%
    #1%
}
\DeclarePairedDelimiterX\softparenthesis[1]{(}{)}
{%
    #1%
}
\NewDocumentCommand{\E}{o}{\mathbb{E}\IfValueT{#1}{_{#1}}}
\NewDocumentCommand{\EX}{soO{}m}{\E[#2]\mathopen{}\IfBooleanTF{#1}{\gullwingsparenthesis*{#4}}{\gullwingsparenthesis[#3]{#4}}}

\makeatletter
\DeclareMathOperator{\Var}{\text{Var}} %
\DeclareMathOperator{\Cov}{\text{Cov}} %
\NewDocumentCommand{\VarX}{soO{}m}{\Var\IfValueT{#2}{_{#2}}\mathopen{}\IfBooleanTF{#1}{\gullwingsparenthesis*{#4}}{\gullwingsparenthesis[#3]{#4}}}
\NewDocumentCommand{\CovX}{soO{}mm}{\Cov\IfValueT{#2}{_{#2}}\mathopen{}\IfBooleanTF{#1}{\gullwingsparenthesis*{#4,\mathopen{}#5}}{\gullwingsparenthesis[#3]{#4,\mathopen{}#5}}}

\DeclarePairedDelimiter\abs{\vert}{\vert}

\DeclarePairedDelimiterX\scalarprod[2]{\langle}{\rangle}{#1,\mathopen{}#2}
\DeclarePairedDelimiter\ceil{\lceil}{\rceil}

\DeclarePairedDelimiterXPP{\sign}[1]{\mathop{sign}\mathopen{}}{(}{)}{}{#1}

\DeclarePairedDelimiterXPP{\ordo}[1]{\mathcal{O}\mathopen{}}{(}{)}{}{#1}

\NewDocumentCommand\set{O{}m}{\gullwingsparenthesis[#1]{#2}}

\NewDocumentCommand\range{O{}m}{\squareparenthesis[#1]{#2}}

\DeclareMathOperator{\Entropy}{\textit{H}}
\newcommand{\HX}[2][]{{\Entropy\subscriptifnonempty{#1}\left\{ #2 \right\}}}
\WithSuffix\newcommand\HX*[2][]{{\Entropy\subscriptifnonempty{#1}\{ #2 \}}}

\newcommand{\minX}[1]{\ensuremath{\min\set{#1}}}

\DeclareMathOperator*{\argmin}{\arg\!\min}

\newcommand{\vect}[1]{{\bm{#1}}}

\DeclareMathOperator{\diag}{diag}
\DeclareMathOperator{\trace}{Tr}
\makeatletter
\newcommand{\traceX}[1]{\trace\parenthesis@round{#1} }
\WithSuffix\newcommand\traceX*[1]{\trace\parenthesis@round@star{#1}}

\DeclareMathOperator{\rank}{rank}
\DeclareMathOperator{\erank}{erank}
\NewDocumentCommand{\rankX}{!s!D(){\cdot}}
{\rank\IfBooleanTF{#1}%
    {\parenthesis@round@star{#2}}%
    {\parenthesis@round{#2}}%
}
\NewDocumentCommand{\erankX}{!s!D(){\cdot}}
{\erank\IfBooleanTF{#1}%
    {\parenthesis@round@star{#2}}%
    {\parenthesis@round{#2}}%
}
\makeatother

\newcommand{\transpose}{^{\mkern-1.5mu\mathsf{T}}}

\newcommand{\identity}[1]{\bm{I}\subscriptifnonempty{#1}}
\newcommand{\onevec}[1]{\mathbf{1}\subscriptifnonempty{#1}}
\newcommand{\zerovec}[1]{\mathbf{0}\subscriptifnonempty{#1}}

\DeclarePairedDelimiter{\norm}{\lVert}{\rVert}

\newcommand{\mnorm}[2][]{
		\left\vert\kern-0.25ex\left\vert\kern-0.25ex\left\vert
		#2
		\right\vert\kern-0.25ex\right\vert\kern-0.25ex\right\vert\subscriptifnonempty{#1}
}

\NewDocumentCommand{\caligraphiccommand}{mmom}{#1\mathopen{}%
    \IfBooleanTF{#2}%
    {\softparenthesis*{\ifstrempty{#4}{\cdot}{#4}}}%
    {\softparenthesis[#3]{\ifstrempty{#4}{\cdot}{#4}}}%
}

\newcommand{\cC}{\mathcal{C}}
\NewDocumentCommand{\cCX}{sO{}m}{\caligraphiccommand{\cC}{#1}[#2]{#3}}

\newcommand{\cD}{\mathcal{D}}

\newcommand{\cG}{\mathcal{G}}

\newcommand{\cQ}{\mathcal{Q}}
\NewDocumentCommand{\cQX}{sO{}m}{\caligraphiccommand{\cQ}{#1}[#2]{#3}}

\newcommand{\cS}{\mathcal{S}}

\newcommand{\reals}{\mathbb{R}}

\def\shrug{\texttt{\raisebox{0.75em}{\char`\_}\char`\\\char`\_\kern-0.5ex(\kern-0.25ex\raisebox{0.25ex}{\rotatebox{45}{\raisebox{-.75ex}"\kern-1.5ex\rotatebox{-90})}}\kern-0.5ex)\kern-0.5ex\char`\_/\raisebox{0.75em}{\char`\_}}}

\usepackage{tabularx}
\usepackage{graphicx}
\PassOptionsToPackage{table}{xcolor}
\PassOptionsToPackage{dvipsnames}{xcolor}
\usepackage{xcolor}

\usepackage{pgf}
\usepackage{pgfplots}
\usepackage{tikz}
\usepackage{tikz-3dplot}

\usetikzlibrary{positioning}
\usetikzlibrary{decorations.pathreplacing}
\usetikzlibrary{plotmarks}
\usetikzlibrary{patterns}
\usetikzlibrary{calc}
\usetikzlibrary{spy}
\usetikzlibrary{fit}
\usetikzlibrary{backgrounds}
\usepgfplotslibrary{groupplots}
\usepgfplotslibrary{fillbetween}
\pgfplotsset{
	compat=1.18,
}
\usepackage{pgfplotstable}

\pgfplotsset{filter discard warning=false}
\pgfplotsset{every axis plot/.append style={line width = 2pt},axis background/.style={fill=white}}

\usetikzlibrary{external}
\usepgfplotslibrary{external}

\makeatletter
\DeclareRobustCommand{\rvdots}{%
	\vbox{
		\baselineskip5\p@\lineskiplimit\z@
		\kern-\p@
		\hbox{.}\hbox{.}\hbox{.}
}}
\makeatother

\tikzset{syscircle/.style={
		draw,
		circle,
		minimum size=0.6cm,
		fill=blue!05},
}
\tikzset{sysblock/.style={
		draw,
		minimum width=1.1cm,
		minimum height=1.0cm,
		fill=red!05},
}
\tikzset{sysline/.style={
	->,
	thick
	},
}
\tikzset{syslinerev/.style={
		sysline,
		<-,
	},
}
\tikzset{sysjunction/.style={
		circle,
		minimum size=1.5mm,
		inner sep=0,
		fill=black
	},
}
\tikzset{sysfitblock/.style={
		inner sep=5mm,
		fill=cyan!02.5,
		densely dashed,
		thick,
		draw=black,
		rounded corners=8pt,
	},
}
\tikzset
{
	port/.style     = {inner sep=0pt, font=\tiny},
	cross/.style =
	{%
		path picture=%
		{
			\draw
			(path picture bounding box.north west) --
			(path picture bounding box.south east)
			(path picture bounding box.south west) --
			(path picture bounding box.north east)
			;
		}
	},
	syssum/.style n args = {4}%
	{%
		syscircle, node distance = 2cm, minimum size=9mm, alias=sum,%
		append after command=%
		{%
			node at (sum.north) [port, below=1pt] {$#1$}
			node at (sum.west) [port, right=1pt] {$#2$}
			node at (sum.south) [port, above=1pt] {$#3$}
			node at (sum.east) [port, left=1pt] {$#4$}
			node at (sum) [port, font=\small] {$\sum$}
		},
	},
}

\tikzset{wireless/.pic=
	{
		\node {\small Placeholder};
	}
}

\tikzset{sysswitch/.pic=
	{
		\node[sysjunction,alias=tttt] at (-4mm,0) {};
        \node[sysjunction] at (4mm,0) {};
        \draw[sysline,-] (tttt) to (3.5mm,3mm);
        \path (-4mm,0) to (3.5mm,-3mm);
	}
}

\tikzset{
	pics/portal/.style args={#1/#2}{
		code={
			  \path[fill=#2,line width=0.0cm] (0.01, -1.39).. controls (0.05, -1.37)
			and (0.1, -1.34) .. (0.07, -1.35).. controls (0.03, -1.36) and (-0.03, -1.37)
			.. (-0.09, -1.36).. controls (-0.27, -1.35) and (-0.42, -1.22) .. (-0.53,
			-0.99).. controls (-0.59, -0.85) and (-0.64, -0.68) .. (-0.63, -0.63)..
			controls (-0.63, -0.62) and (-0.62, -0.62) .. (-0.62, -0.62).. controls
			(-0.62, -0.62) and (-0.6, -0.67) .. (-0.58, -0.72).. controls (-0.47, -1.02)
			and (-0.28, -1.21) .. (-0.08, -1.21).. controls (0.08, -1.21) and (0.29,
			-1.07) .. (0.43, -0.86).. controls (0.59, -0.63) and (0.7, -0.34) .. (0.75,
			-0.06).. controls (0.78, 0.1) and (0.79, 0.17) .. (0.79, 0.34).. controls
			(0.79, 0.55) and (0.77, 0.71) .. (0.72, 0.89) -- (0.7, 0.96) -- (0.71, 0.75)..
			controls (0.71, 0.44) and (0.7, 0.28) .. (0.64, 0.13).. controls (0.62, 0.07)
			and (0.58, 0.01) .. (0.58, 0.02).. controls (0.57, 0.03) and (0.58, 0.08) ..
			(0.59, 0.15).. controls (0.61, 0.3) and (0.62, 0.51) .. (0.61, 0.65)..
			controls (0.59, 1.02) and (0.47, 1.25) .. (0.26, 1.31).. controls (0.15, 1.35)
			and (0.08, 1.34) .. (-0.04, 1.28).. controls (-0.23, 1.18) and (-0.38, 1.02)
			.. (-0.5, 0.78).. controls (-0.61, 0.56) and (-0.69, 0.27) .. (-0.75, -0.13)..
			controls (-0.78, -0.31) and (-0.78, -0.55) .. (-0.76, -0.7).. controls
			(-0.73, -0.92) and (-0.66, -1.09) .. (-0.55, -1.21).. controls (-0.48, -1.3)
			and (-0.36, -1.37) .. (-0.25, -1.4).. controls (-0.16, -1.42) and (-0.05,
			-1.42) .. (0.01, -1.39) -- cycle;

			\path[fill=#1,line width=0.0cm] (-0.3, -1.59).. controls (-0.44, -1.57)
			and (-0.53, -1.52) .. (-0.64, -1.42).. controls (-0.86, -1.19) and (-0.98,
			-0.88) .. (-1.0, -0.46).. controls (-1.01, -0.25) and (-0.97, 0.15) .. (-0.91,
			0.42).. controls (-0.84, 0.78) and (-0.69, 1.11) .. (-0.54, 1.29).. controls
			(-0.5, 1.33) and (-0.45, 1.38) .. (-0.44, 1.37).. controls (-0.44, 1.37) and
			(-0.45, 1.34) .. (-0.47, 1.3).. controls (-0.58, 1.13) and (-0.72, 0.82) ..
			(-0.72, 0.77).. controls (-0.72, 0.75) and (-0.72, 0.75) .. (-0.71, 0.77)..
			controls (-0.55, 1.07) and (-0.36, 1.32) .. (-0.17, 1.45).. controls (-0.1,
			1.5) and (0.02, 1.57) .. (0.09, 1.58).. controls (0.32, 1.64) and (0.59, 1.54)
			.. (0.76, 1.34).. controls (0.88, 1.2) and (0.96, 1.0) .. (0.99, 0.76)..
			controls (1.0, 0.67) and (1.0, 0.34) .. (0.99, 0.23).. controls (0.96, -0.05)
			and (0.84, -0.47) .. (0.71, -0.78).. controls (0.5, -1.25) and (0.24, -1.53)
			.. (-0.06, -1.59).. controls (-0.12, -1.6) and (-0.24, -1.6) .. (-0.3, -1.59)
			-- cycle(0.01, -1.39).. controls (0.05, -1.37) and (0.1, -1.34) .. (0.07,
			-1.35).. controls (0.03, -1.36) and (-0.03, -1.37) .. (-0.09, -1.36)..
			controls (-0.27, -1.35) and (-0.42, -1.22) .. (-0.53, -0.99).. controls
			(-0.59, -0.85) and (-0.64, -0.68) .. (-0.63, -0.63).. controls (-0.63, -0.62)
			and (-0.62, -0.62) .. (-0.62, -0.62).. controls (-0.62, -0.62) and (-0.6,
			-0.67) .. (-0.58, -0.72).. controls (-0.47, -1.02) and (-0.28, -1.21) ..
			(-0.08, -1.21).. controls (0.08, -1.21) and (0.29, -1.07) .. (0.43, -0.86)..
			controls (0.59, -0.63) and (0.7, -0.34) .. (0.75, -0.06).. controls (0.78,
			0.1) and (0.79, 0.17) .. (0.79, 0.34).. controls (0.79, 0.55) and (0.77, 0.71)
			.. (0.72, 0.89) -- (0.7, 0.96) -- (0.71, 0.75).. controls (0.71, 0.44) and
			(0.7, 0.28) .. (0.64, 0.13).. controls (0.62, 0.07) and (0.58, 0.01) .. (0.58,
			0.02).. controls (0.57, 0.03) and (0.58, 0.08) .. (0.59, 0.15).. controls
			(0.61, 0.3) and (0.62, 0.51) .. (0.61, 0.65).. controls (0.59, 1.02) and
			(0.47, 1.25) .. (0.26, 1.31).. controls (0.15, 1.35) and (0.08, 1.34) ..
			(-0.04, 1.28).. controls (-0.23, 1.18) and (-0.38, 1.02) .. (-0.5, 0.78)..
			controls (-0.61, 0.56) and (-0.69, 0.27) .. (-0.75, -0.13).. controls (-0.78,
			-0.31) and (-0.78, -0.55) .. (-0.76, -0.7).. controls (-0.73, -0.92) and
			(-0.66, -1.09) .. (-0.55, -1.21).. controls (-0.48, -1.3) and (-0.36, -1.37)
			.. (-0.25, -1.4).. controls (-0.16, -1.42) and (-0.05, -1.42) .. (0.01, -1.39)
			-- cycle;

		}
	}
}

\pgfplotscreateplotcyclelist{cyclelist}{
	blue, solid, every mark/.append style={fill=blue}, mark=*,mark repeat={10},mark size=1pt\\
	red,  solid, every mark/.append style={fill=red}, mark=triangle*,mark repeat={10},mark size=1pt\\
	cyan, dashdotdotted \\
	red, densely dashed,every mark/.append style={fill=red}, mark=diamond*,mark repeat={10},mark size=1pt, mark phase = {5},\\
	darkgray, densely dotted\\
	orange, dashdotted\\
}

\pgfplotsset{singlecolumnsize/.style={
  width = 8.6cm,
  height = 4.8 cm,
  },
}

\let\hat\widehat
\let\tilde\widetilde
\let\check\widecheck

\usepackage[noabbrev,capitalise]{cleveref} %
\mathpreamblepostcleveref

\loadpreambletheoremsI

\usepackage{suffix}
\newcommand{\ts}[1]{^{(#1)}} %
\WithSuffix\newcommand\ts*[1]{{\ts{#1}}} %

\NewDocumentCommand{\MeasureCorrSVD}{m}{\textnormal{MCorrSVD}\mathopen{}(#1)}
\NewDocumentCommand{\MeasureCorrPCA}{m}{\textnormal{MCorrPCA}\mathopen{}(#1)}
\NewDocumentCommand{\NarrowPCA}{m}{\textnormal{TruncPCA}\mathopen{}(#1)}
\NewDocumentCommand{\NarrowSVD}{m}{\textnormal{TruncSVD}\mathopen{}(#1)}
\NewDocumentCommand{\PCAFedUpdate}{m}{\textnormal{PCAFedStateUpdate}\mathopen{}(#1)}

\DeclareMathOperator{\css}{CSS}
\NewDocumentCommand\cssX{smm}{\css\IfBooleanTF{#1}{\softparenthesis*{#2,\mathopen{}#3}}{\softparenthesis{#2,\mathopen{}#3}}}

\newcommand\smallnegvspace{}
\title{Exploiting Correlations in Federated Learning: Opportunities and Practical Limitations
}

\author{
    Adrian Edin,~\IEEEmembership{Student Member,~IEEE,} Michel Kieffer,~\IEEEmembership{Senior Member,~IEEE,} \\Mikael Johansson,~\IEEEmembership{IEEE Fellow}, and Zheng Chen,~\IEEEmembership{Senior Member,~IEEE}
\thanks{
This work was supported in part by ELLIIT, the Swedish Research Council (VR), the Knut and Alice Wallenberg (KAW) Foundation, and the Wallenberg AI, Autonomous Systems and Software Program (WASP) funded by the Knut and Alice Wallenberg Foundation.
}
\thanks{Adrian Edin and Zheng Chen are with the Department of Electrical Engineering (ISY), Linköping University, SE-581 83 Linköping, Sweden (email: \{adrian.edin, zheng.chen\}@liu.se).}
\thanks{Michel Kieffer is with CentraleSupélec, CNRS, L2S, Université Paris-Saclay, Gif-sur-Yvette, FR-91192, France (email: michel.kieffer@l2s.centralesupelec.fr).}
\thanks{Mikael Johansson is with the Division of Decision and Control Systems, School of EECS, KTH Royal Institute of Technology, SE-100 44 Stockholm, Sweden (email: mikaelj@kth.se).}
}

\begin{document}

\maketitle

\begin{abstract}
The communication bottleneck in federated learning (FL) has spurred extensive research into techniques to reduce the volume of data exchanged between client devices and the central parameter server. In this paper, we systematically classify gradient and model compression schemes into three categories based on the type of correlations they exploit: \textit{structural}, \textit{temporal}, and \textit{spatial}. We examine the sources of such correlations, propose quantitative metrics for measuring their magnitude, and reinterpret existing compression methods through this unified correlation-based framework.
Our experimental studies demonstrate that the degrees of structural, temporal, and spatial correlations vary significantly depending on task complexity, model architecture, and algorithmic configurations.
These findings suggest that algorithm designers should carefully evaluate correlation assumptions under specific deployment scenarios rather than assuming that they are always present.
Motivated by these findings, we propose two adaptive compression designs that actively switch between different compression modes based on the measured correlation strength, and we evaluate their performance gains relative to conventional non-adaptive approaches. In summary, our unified taxonomy provides a clean and principled foundation for developing more effective and application-specific compression techniques for FL systems.
\end{abstract}


\acronymslinktofirstinstance

\section{Introduction}\label{sec:introduction}
\smallnegvspace

As \gls{ml} models grow in scale and complexity, training them on massive datasets with a single machine becomes increasingly demanding. The rise of edge computing and data parallelism has driven the emergence of collaborative machine learning, where multiple nodes—each with access to a portion of the data—collaborate to train a shared model without sharing raw samples. This can be done in a network with a master–worker architecture, such as \gls{fl}, or in a decentralized topology with local interactions, such as decentralized learning.  In server-based \gls{fl}, multiple clients iteratively update their local copy of the global model using their own private data. These model updates are then sent to a central \gls{ps}, which aggregates them to obtain an updated global model. Compared to centralized learning, \gls{fl} offers advantages such as improved scalability and enhanced data privacy. However, \gls{fl} also faces the challenge of a communication bottleneck due to frequent information exchange between local clients and the \gls{ps}, especially when the exchanged information contains high-dimensional parameter vectors from deep neural networks.

To address this communication bottleneck, several strategies for improving the communication efficiency of \gls{fl} have been proposed based on
\begin{itemize}
    \item reducing the number of bits transmitted from clients to the server in each communication round, \textit{e.g.}, through \emph{model update compression}~\cite{konevcny2016federated};
    \item reducing the number of active communication links, \textit{e.g.}, via client selection~\cite{cho2022towards,ko2023selectionbandwidth} or  event-triggered communication~\cite{chen2018LAG,xu2024selectioncompression};
    \item reducing the frequency of communication, \textit{e.g.}, by performing multiple gradient steps before aggregation~\cite{stich2019local}.
\end{itemize}
In earlier studies, commonly used compression techniques in \gls{fl}, such as quantization~\cite{alistarh2017QSGD} and random sparsification~\cite{wangni2018GradientSparsification}, treat the model updates generated by local clients as independent information sources. However, the correlation and redundancy among local model updates can be exploited to achieve more efficient compression. These correlations may arise from the model architecture, the learning algorithm, or other design choices in the \gls{fl} system ~\cite{wang2018ATOMO, wang2023SVDFed,azam2022RecyclingModelUpdates,richtarik2021EF21}. Classical coding theory shows that correlated sources can be compressed more efficiently, with greater savings as correlation increases~\cite{sayood2018}. For instance, in the lossy compression of parallel Gaussian sources, correlation enables lower transmission rates for a given distortion level compared to uncorrelated sources. Another example is predictive coding, widely used in image and video compression, that exploits redundancy between adjacent pixels or consecutive frames to enhance compression efficiency. Inspired by this principle, recent works in \gls{fl}~\cite{yue2022PredictiveCoding, edin2024TemporalPredictiveCoding,adikari2021CompressingTemporalCorrelation} apply predictive techniques to compress temporally correlated model updates. By forecasting future updates from past ones, only the residuals—representing deviations from the prediction—need to be transmitted, significantly reducing communication overhead when consecutive updates are sufficiently correlated.
So far, predictive coding has mainly been applied to exploit temporal correlations in local model updates, but the underlying principle is not limited to the time domain. In fact, similar strategies could also be applied to capture structural correlations within the model itself or spatial correlations across clients, for example, using subspace projection methods~\cite{park2023regulatedsubspace,zhang2025efficient,han2024randomprojection}. However, whether such correlations are present in practical \gls{fl} settings and how they might be effectively measured and leveraged for compression remains an open question.
Given the growing interest in redundancy-based compression schemes %
in the \gls{fl} literature, we find it pertinent to examine \textit{whether such correlations are observable} in practice, and \textit{to what extent they can actually contribute} to improving communication efficiency.

Our paper makes the following contributions:
\begin{itemize}
    \item We introduce a unified framework for analyzing \textit{structural}, \textit{temporal}, and \textit{spatial} correlations in \gls{fl}, and propose metrics for measuring each type of correlation.
    \item We reinterpret existing redundancy-based compression techniques through the lens of these correlations, offering a new perspective on their underlying design principles.
    \item The practical limitations of correlation-based subspace projection methods, such as computational complexity and memory overhead, are also discussed to provide a thorough assessment of their advantages and disadvantages.
    \item Based on existing non-adaptive compression schemes, we propose two adaptive designs that measure the strength of correlations on the fly and dynamically switch between different compression modes.
    \item We conduct experiments demonstrating that correlation strength varies with model size, task complexity, and system configuration. These results highlight the importance of adaptive compression designs compared to their static counterparts.
\end{itemize}

\subsubsection*{Notation}
The set of real numbers is denoted by $\reals$.
Vectors are written in bold lowercase letters, \textit{e.g.}, $\vect{a}$, and matrices in uppercase letters, \textit{e.g.}, $A$.
The symbols $\norm{\cdot}$ and $\norm{\cdot}_F$ denote the Euclidean norm and Frobenius norm, respectively.
For a set $\cS$, $\abs{\cS}$ denotes its cardinality.
For a positive integer $N$, $[N]$ denotes the set $\set{1,\ldots,N}$.

\section{Three Sources of Correlation in Federated Learning}\label{sec:correlation}
\smallnegvspace

In a \gls{fl} system, a set of $K$ distributed clients collaborates in training a shared \gls{ml} model under the coordination of a central \gls{ps}. Each client $k \in [K]$ has local access to subset $\cD_k$ of the complete dataset $\cD=\cup_{k=1}^K \cD_k$. The training objective is to minimize a global loss
\begin{align}
    f(\vect{x}) = \sum_{k=1}^K w_k f_k(\vect{x}),
\end{align}
where $\vect{x}\in \reals^d$ is the vector of model parameters, $w_k = \vert \cD_k\vert/\vert \cD\vert$ are aggregation weights proportional to the amount of data held by client $k$,  $f_k(\vect{x}) = \frac{1}{|\cD_k|}\sum_{\xi \in \cD_k} \ell(\vect{x}; \xi)$ is the local objective of client $k$, and $\ell(\vect{x}; \xi)$ is the loss for sample $\xi$.
A canonical \gls{fl} algorithm is \gls{fedavg}~\cite{mcmahan2017communication}, which proceeds iteratively as follows:
\begin{enumerate}
    \item The \gls{ps} broadcasts the current global model parameters $\vect{x}\ts{t}$ to all clients.
    \item Each client $k$ performs $\tau$ local stochastic gradient updates using mini-batches
    \[
    \mkern -3mu \vect{x}_k\ts{t,i+1} = \vect{x}_k\ts{t,i} - \gamma \nabla f_k(\vect{x}_k\ts{t,i}, \cD_k\ts{t,i}), \ i = 0, \ldots, \tau-1,
    \]
    where $\gamma$ is the learning rate, and $\cD_k\ts{t,i}$ is a randomly sampled mini-batch of data from $\cD_k$.
    \item Each client computes its \emph{local model update} $\vect{g}_k\ts{t} = \vect{x}_k\ts{t,\tau} - \vect{x}\ts{t}$ and sends it to the \gls{ps}.\footnote{With single-step SGD, i.e., $\tau=1$, the local model update equals the negative of the local gradient multiplied by the step size.}
    \item The \gls{ps} aggregates all updates to obtain  the new global model:
    \begin{align}\label{eq:gd step}
        \vect{x}\ts{t+1} = \vect{x}\ts{t} + \sum_{k=1}^{K} w_k \vect{g}_k\ts{t}.
    \end{align}
\end{enumerate}

Many variations of \gls{fedavg} have been proposed, such as reducing variance by computing and communicating auxiliary correction vectors~\cite{karimireddy2020scaffold,zhao2024fasterrates}, introducing momentum in local updates~\cite{liu2020accelerating}, or adding a proximal term to improve stability in heterogeneous settings~\cite{li2020federated}.

\paragraph{Communication Bottleneck} In each round of the distributed training procedure, \gls{fedavg} requires two communication phases: (1) clients upload local updates to the \gls{ps}; and (2) the \gls{ps} broadcasts the updated global model.
Most existing research on communication-efficient \gls{fl} focuses on client-to-\gls{ps} links, since each client requires dedicated resources to transmit its model update to the \gls{ps}. To reduce the communication cost, data compression is commonly applied before transmitting the local model updates.
Let $\{\hat{\vect{g}}_k\ts{t}\}_{k=1,\ldots,K}$ represent the reconstructed local model updates at the \gls{ps}, a common design principle for compression is to minimize the distortion measured by the empirical \gls{mse}, given by
\begin{equation}
    \textnormal{MSE}(\hat{\vect{g}}_k\ts{t}, \vect{g}_k\ts{t})=\frac{1}{d}\norm{\hat{\vect{g}}_k\ts{t}-\vect{g}_k\ts{t}}^2.
\end{equation}
Many existing works use this distortion measure implicitly or explicitly, as we will discuss later.

Earlier studies on \gls{fl} with compressed updates focus on simple schemes such as sparsification or masking. Recent studies have adopted different forms of subspace projection methods, motivated by the potential correlation/redundancy in transmitted information~\cite{wang2023SVDFed,adikari2021CompressingTemporalCorrelation,xue2022FedOComp,yu2018GradiVeQ,abrahamyan2022LearnedGradientCompression,azam2022RecyclingModelUpdates}.
Nevertheless, current literature lacks a unified taxonomy to describe and interpret how specific compression designs relate to different sources of correlation inherent in \gls{fl} algorithms. We categorize them into three types:
\begin{itemize}
    \item \textit{Structural correlation} results from the learning model, \emph{e.g.}, neural network structure. Some components of the gradient vectors may be correlated with other components.
    \item \textit{Temporal correlation} originates from the iterative nature of the optimization algorithm and is influenced by design features such as momentum and regularization. As a result, gradient vectors generated in consecutive iterations may exhibit significant correlation.
    \item \textit{Spatial correlation} arises when training datasets across different clients share similar distributions. This can lead to correlated local model updates across clients.
\end{itemize}
In the following sections, we discuss each correlation type in detail and provide illustrative examples. While a complete theoretical understanding of these correlations remains elusive -- partly due to the black-box nature of deep learning models~\cite{zhang2017UnderstandingDeepLearning} -- we aim to offer practical insight and useful tools for analyzing and exploiting them.

\section{Structural Correlation}\label{sec:correlation:structural}
\smallnegvspace
We define structural correlation as the dependency among the elements within a client's local model update $\vect{g}_k\ts{t}$.
Let us consider a simple linear regression example. Given a dataset $\cD$ containg $N$ pairs of feature-observation points $(\vect{\xi}_i,y_i)\in\reals^{d-1}\times\reals$, $i\in[N]$, we form a prediction function $\hat{y}_i$ using a linear model
\begin{align}
    \hat{y_i}(\vect{w},b) = \vect{\xi}_i\transpose\vect{w}+b,
\end{align}
where $\vect{w}\in\reals^{d-1}$ is the weight vector and $b\in\mathbb{R}$ is the bias.
The objective is to find the estimate of $\vect{w}$ and $b$ such that the corresponding elements of $\hat{\vect{y}} = (\hat{y}_1,\dots,\hat{y}_N)\transpose\in\reals^N$ and $\vect{y}= (y_1,\dots,y_N)\transpose\in\reals^N$ are as close as possible for all samples $i\in\cD$. Using vector notation, we obtain $\hat{\vect{y}}(\vect{x})=R\vect{x}$, where $R=[\vect{\xi}_1,\ldots, \vect{\xi}_N; \onevec{N}\transpose]\transpose \in \reals^{N\times d}$, and $\vect{x}=[\vect{w};b]\in\reals^{d}$ collects all tunable parameters.

We quantify the quality of the fit using the \gls{mse} over all $N$ samples. The objective and its gradient are
\begin{align}\label{eq:toy:objective}
    f(\vect{x}) = \frac{\norm{\hat{\vect{y}}(\vect{x}) - \vect{y}}^2}{N} = \frac{\vect{x}\transpose R\transpose R\vect{x} - 2\vect{y}\transpose R \vect{x}  + \vect{y}\transpose\vect{y}}{N}
\end{align}
and
\begin{align}\label{eq:toy:gradient}
    \nabla f(\vect{x}) = \frac{1}{N} \left( 2R\transpose R\vect{x}-2R\transpose\vect{y}\right),
\end{align}
respectively.
These expressions display how the dataset $R$ and the objective function $f(\vect{x})$ interact to potentially induce correlations among the components of $\nabla f(\vect{x})$.

For more complex tasks with larger \gls{ml} models, several existing works have investigated the existence of such structural correlation.
For instance,
\cite{gur-ari2018TinySubspace} and~\cite{cosson2023LowRankGradientDescent} show that the gradient converges to a very small subspace spanned by a few top eigenvectors of the Hessian. This suggests that dimension reduction can be achieved by projecting the gradients onto a low-dimensional subspace~\cite{li2021LowDimensionalLandscape}. The impact of the network model and data on the spectrum of the Hessian matrix has been investigated in ~\cite{sagun2018EmpiricalHessianOverParametrized}. The observations can also be related to the \textit{intrinsic dimensionality} of data and model representations~\cite{ansuini2019IntrinsicRepresentationDimensions, li2018IntrinsicDimension}, suggesting that high-dimensional models with a large number of parameters may effectively be trained within a lower-dimensional subspace. %

\subsection{How to Evaluate the Degree of Structural Correlation?}\label{sec:evaluate correlations}
\smallnegvspace

First, consider the local model update vector $\vect{g}_k\ts{t}\in\reals^d$,
we introduce the \textit{model update matrix} generated by client $k$ at iteration $t$ as
\begin{align}\label{eq:gradient matrix slices}
    G_k\ts{t}=\begin{pmatrix}\vect{g}_{k,1}\ts{t} & \dots & \vect{g}_{k,n}\ts{t}\end{pmatrix}\in\reals^{m\times n},
\end{align}
where $\vect{g}_{k,i}\ts{t}\in\reals^m$ is the $i$-column of $G_k\ts{t}$.
This matrix contains the elements of $\vect{g}_k\ts{t}\in\reals^d$, reshaped into an $m\times n$ matrix such that $mn=d$.\footnote{If $d$ is not divisible by $m$, zero-padding will be applied. } We refer to each column $\vect{g}_{k,i}\ts{t}$ as a \textit{model update slice}.
The structural correlation may depend on the organization of the model update matrix, as evidenced in~\cite{yu2018GradiVeQ, vogels2019PowerSGD,wang2018ATOMO}, where different choices for the mapping from vector $\vect{g}_k\ts{t}$ to matrix $G_k\ts{t}$ have been considered.

The cosine similarity is used to measure the degree of structural correlation between two model update slices:
\begin{equation}
    \css(\vect{g}_{k,i}\ts{t},\vect{g}_{k,j}\ts{t}) \!=\!
    \begin{cases}
        \frac{(\vect{g}_{k,i}\ts{t})\transpose\vect{g}_{k,j}\ts{t}}{\norm{\vect{g}_{k,i}\ts{t}}\norm{\vect{g}_{k,j}\ts{t}}},& \vect{g}_{k,i}\ts{t}\!\neq\!\zerovec{}~\text{and}~\vect{g}_{k,j}\ts{t}\!\neq\!\zerovec{}
        \nonumber\\
        0,&\text{otherwise}
    \end{cases}
\end{equation}
where $i$ and $j$ are slice indices.
In \gls{fl}, if different model update slices of the model update matrix exhibit high pairwise cosine similarity, it becomes possible to identify a common subspace to represent these model update slices effectively.
This enables a more efficient, lower-dimensional representation, which is the main motivation behind many compression designs using low-rank approximation.

Another way to quantify the degree of structural correlation is to use \gls{svd}.
A matrix $G\in\reals^{m\times n}$ can be decomposed as
\begin{align}\label{eq:def:svd}
    G = U \Sigma V\transpose,
\end{align}
where $U\in\reals^{m\times m}$ and $V\in\reals^{n\times n}$ are orthonormal matrices, and $\Sigma\in\reals^{m\times n}$ contains the singular values $\sigma_1,\dots,\sigma_{\minX{m,n}}$ on its main diagonal, in decreasing order $\sigma_1\geq\dots\geq\sigma_{\minX{m,n}}\geq0$.
\gls{svd} enables the identification of global correlations among multiple vectors, rather than local pairwise correlations, albeit at a higher computational cost than the cosine similarity metric.
When applying \gls{svd} to the model update matrix, the presence of a few dominant singular values suggests a stronger low-rank structure.

\subsection{Exploiting Structural Correlation using Low-Rank Approximation} \label{sec:exploiting correlations}
\smallnegvspace
Once a high degree of structural correlation is observed within each local model update, low-rank approximation can be applied to obtain a compressed update of reduced dimension.

\paragraph{\glsxtrshort{svd}}
One example of low-rank approximation is to perform \gls{svd} on the model update matrix $G_k\ts{t}=U \Sigma V\transpose$, where $U\in\reals^{m\times m}$ and $V\in\reals^{n\times n}$ are orthogonal matrices and $\Sigma$ contains the singular values.
A lower-dimensional representation of $G_k\ts{t}$ can be formed using  $U_r\in\reals^{m\times r}$ and $V_r\in \reals^{n\times r}$, whose columns are the first $r$ columns of $U$ and $V$ respectively, \textit{i.e.}, $U_r=[U]_{:,1:r}$ and $V_r=[V]_{:,1:r}$.
Model update compression can use either $U_r$ alone or the combination of $U_r$ and $V_r$.
\begin{itemize}
    \item When compression uses only $U_r$, client $k$ transmits the compressed update $U_r\transpose G_k\ts{t}\in\reals^{r\times n}$ to the \gls{ps}. The communication cost is $rn$ in each client-to-\gls{ps} link. The \gls{ps} must also know $U_r$ to reconstruct the original update, which requires transmitting $rm$ elements.
    \item When compression uses both $U_r$ and $V_r$, client $k$ transmits the diagonal matrix $\tilde{G}_k\ts{t}=U_r\transpose G_k\ts{t} V_r \in\reals^{r\times r}$ to the \gls{ps}. The communication cost is $r$ (the number of diagonal elements in the matrix) in each client-to-\gls{ps} link.
    Additional transmissions of $rm$ elements for $U_r$ and $rn$ elements for $V_r$ are required for the \gls{ps} to reconstruct $U_r \tilde{G}_k\ts{t}V_r\transpose$.
\end{itemize}
An example of using \gls{svd} to compress local model updates is the \textit{ATOMO} framework presented in~\cite{wang2018ATOMO}.
\begin{remark}
    By ranking the singular values in decreasing order, and sending the basis vectors corresponding to the $r$ largest singular values, the compression design effectively minimizes the \gls{mse} distortion of the reconstructed model update, under a limited number of transmitted elements.
\end{remark}

\paragraph{\glsxtrshort{pca}} \gls{pca}~\cite{yu2018GradiVeQ,xue2022FedOComp,li2021LowDimensionalLandscape} is another efficient tool for obtaining an approximate representation of the model update matrix $G_k\ts{t}$ using a set of vectors that capture the largest variations in the columns of $G_k\ts{t}$. This is achieved by computing a covariance matrix $C$ from the columns of $G_k\ts{t}$ and performing the eigenvalue decomposition $C = Q\Lambda Q\transpose$, where $\Lambda =\diag(\lambda_1,\dots,\lambda_m)$ is the diagonal matrix of eigenvalues, with $\lambda_1\geq\dots\geq\lambda_m\geq0$, and $Q\in\reals^{m\times m}$ is an orthonormal matrix. Given the model update matrix  $G_k\ts{t}$ and the mean $\bar{\vect{g}} = \frac{1}{n}\sum_{i'=1}^{n} \vect{g}_{k,i'}\ts{t}\in\reals^{m}$, the covariance matrix is computed as
\begin{equation}
    C = \frac{1}{n-1}\sum_{i'=1}^{n} (\vect{g}_{k,i'}\ts{t}-\bar{\vect{g}})(\vect{g}_{k,i'}\ts{t}-\bar{\vect{g}})\transpose.
\end{equation}

The principal components are the columns of $Q$. Consider the matrix $Q_r=[Q]_{:,1:r}$ of the first $r$ columns of $Q$. To exploit the structural correlation, client $k$ can compute a reduced-dimensional representation of the $i$-th model update slice as
\begin{align}\label{eq:pca:compression}
    \tilde{\vect{g}}_{k,i}\ts{t} = Q_r\transpose (\vect{g}_{k,i}\ts{t} - \bar{\vect{g}})\in\reals^{r},
\end{align}
which is then transmitted to the \gls{ps}.
After the \gls{ps} receives the compressed model update slice from client $k$, it can get an estimate of the model update slice by
\begin{align}\label{eq:pca:decompression}
    \hat{\vect{g}}_{k,i}\ts{t} = Q_r\tilde{\vect{g}}\ts{t} + \bar{\vect{g}} = Q_rQ_r\transpose(\vect{g}_{k,i}\ts{t} - \bar{\vect{g}}) +\bar{\vect{g}}\in\reals^{m}.
\end{align}
While the communication cost in the client-to-\gls{ps} link for each model update slice is reduced to $r$ elements instead of $m$, ensuring that both parties have knowledge of $Q_r$ and $\bar{\vect{g}}$ requires additional transmission of $m(r+1)$ elements for all slices.

\begin{remark}
    The \gls{pca} and \gls{svd} are closely related~\cite{gerbrands1981relationships}.
    For example, when $\bar{\vect{g}} = \zerovec{m}$, the squared singular values $\sigma^2_i$ coincide with the eigenvalues of the covariance matrix $\lambda_i$, and the left unitary matrix $U$ coincides with the \gls{pca} basis $Q$.
\end{remark}

For both \gls{pca} and \gls{svd}, it is obvious that the choice of $r$ (the number of basis vectors to be communicated) plays a critical role. Larger values of $r$ capture more information in the subspace spanned by the model update matrix, while smaller values of $r$ reduce the communication cost.
There is always a trade-off among information loss, computational complexity, and communication cost.

\begin{figure}[t!]
    \centering
    \includegraphics[width=0.9\linewidth]{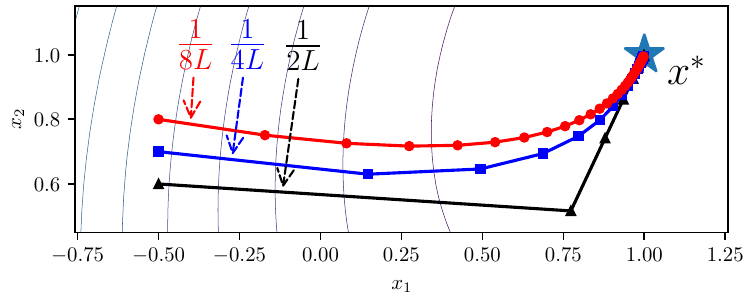}
    \caption{A 2D toy example with gradient descent and the quadratic loss function defined in \eqref{eq:toy:objective}. The trajectories marked with different colors correspond to different values of the learning rate $\gamma \in \{\frac{1}{2L}, \frac{1}{4L}, \frac{1}{8L}\}$.}%
    \label{fig:temporal correlation}
\end{figure}
\section{Temporal Correlation}\label{sec:correlation:temporal}
In the context of \gls{fl}, temporal correlation refers to the similarity between consecutive local model updates $\vect{g}_k\ts{t}, \vect{g}_k\ts{t+1},\ldots$, generated by each client $k$.
In general, the degree of temporal correlation may be affected by the following factors:
\begin{itemize}
    \item Properties of the loss function, \textit{e.g.}, convexity, and smoothness. With a convex and smooth global objective function, we expect stronger temporal correlation compared to a non-convex function with many local minima.
    \item Hyperparameter values in the iterative optimization algorithm, \emph{e.g.}, learning rate, batch size, momentum, and regularization factor.
\end{itemize}
\cref{fig:temporal correlation} shows the trajectory of the parameter vector (with two parameters) in a linear regression problem. We observe that a smaller learning rate yields a smoother trajectory, implying a higher degree of temporal correlation, albeit at the cost of requiring more iterations to converge.

\subsection{How to Evaluate the Degree of Temporal Correlation?}
\smallnegvspace
The degree of temporal correlation can be quantified using the cosine similarity between two model update vectors, \textit{i.e.}, $\css(\vect{g}_{k}\ts{t_1},\vect{g}_{k}\ts{t_2})$,
where $t_1$ and $t_2$ are two time (iteration) indices.
A high degree of similarity suggests that past model updates can be leveraged to predict the trajectory of future updates, which is the main motivation behind prediction-based model update compression schemes.

For the linear regression problem with a quadratic loss function, we can obtain a closed-form expression for the degree of temporal correlation. Combining the gradient vector in \eqref{eq:toy:gradient} with the parameter update $\vect{x}\ts{t+1} = \vect{x}\ts{t} - \gamma \nabla f(\vect{x}\ts{t})$, we get
\begin{align}\label{eq:toy:temporal}
    \nabla f(\vect{x}\ts{t+1}) =& \frac{1}{N}\left( 2R\transpose R\left(\vect{x}\ts{t} - \gamma\nabla f(\vect{x}\ts{t}) \right) - 2R\transpose\vect{y} \right)
    \nonumber\\
    =&  \nabla f(\vect{x}\ts{t}) -\gamma\frac{2R\transpose R}{N} \nabla f(\vect{x}\ts{t})
    \nonumber\\
    =& \left( \identity{}-\gamma\frac{2R\transpose R}{N} \right) \nabla f(\vect{x}\ts{t}).
\end{align}
Define $ A = \identity{}-\gamma\frac{2R\transpose R}{N}$, and let $L=\lambda_\textrm{max}(\frac{R\transpose R}{N})$, where $\lambda_\textrm{max}(\cdot)$ extracts the maximum eigenvalue. Using a step size $0\leq\gamma \leq \frac{1}{L}$ ensures that $\norm{A}\leq 1$, which in turn implies $\norm{\nabla f(\vect{x}\ts{t+1})} \leq \norm{\nabla f(\vect{x}\ts{t})}$ in \eqref{eq:toy:temporal}.
The cosine similarity between two consecutive gradients is thus lower-bounded by
\begin{align}
    \css(\nabla f(&\vect{x}\ts{t+1}),\nabla f(\vect{x}\ts{t}))
    =  \frac{\nabla f(\vect{x}\ts{t+1})\transpose \nabla f(\vect{x}\ts{t})}{\norm{\nabla f(\vect{x}\ts{t+1})}\norm{\nabla f(\vect{x}\ts{t})}}
    \nonumber \\
    \geq & \frac{\nabla f(\vect{x}\ts{t}) \transpose A \nabla f(\vect{x}\ts{t})}    {\norm{\nabla f(\vect{x}\ts{t})}^2 }
    \geq \min_i \lambda_{i}(A),
\end{align}
where $\lambda_{i}(A)$ is the $i$-th eigenvalue of $A$.
Because $\lambda_i(A) = 1-2\gamma\lambda_i(\frac{R\transpose R}{N})$, using $\gamma\leq \frac{1}{2L}$ gives $\min_i \lambda_{i}(A) = \lambda_\textrm{min}(A) = 1-2\gamma L\geq 0$.
Naturally, this means that a smaller learning rate and a smaller $L$ lead to a higher degree of temporal correlation.

\subsection{Exploiting Temporal Correlation using Predictive Coding}
\label{sec:correlation:temporal:exploit}
\smallnegvspace
Predictive coding techniques are widely used in multimedia compression scenarios to reduce redundancy by exploiting correlations between adjacent pixels or image frames~\cite{sayood2018,salomon2010Handbook}.
In \gls{fl}, prediction-based encoding has been used to exploit the temporal correlation between consecutive model updates~\cite{yue2022PredictiveCoding,mishchenki2024CompressedGradientDifferences,richtarik2021EF21, adikari2021CompressingTemporalCorrelation,chen2022SparsifiedGradientDifferences,edin2024TemporalPredictiveCoding}.
The key idea is to generate a prediction $\check{\vect{g}}_k\ts{t}$ of the current model update based on the knowledge of previous updates $\{\vect{g}_k\ts{t-1}, \vect{g}_k\ts{t-2}, \ldots\}$ and transmit only the prediction residual $\vect{g}_k\ts{t}-\check{\vect{g}}_k\ts{t}$, \textit{i.e.}, the difference between the actual update and the prediction. The \gls{ps} computes the same prediction using shared information between the clients, based on which it can reconstruct the original model update after receiving the residual. When the prediction is accurate, the residual contains much less information than the original model update, and can therefore be represented using fewer bits. Communication efficiency can be further improved by omitting residual transmission when the prediction is sufficiently accurate~\cite{edin2024TemporalPredictiveCoding,azam2022RecyclingModelUpdates}. The predictor can, \textit{e.g.},
\begin{itemize}
    \item the previous local model update (\textit{e.g.}, gradient vector when using single-step SGD)~\cite{mishchenki2024CompressedGradientDifferences,richtarik2021EF21}, so that the prediction residual becomes the model update difference;
    \item momentum-inspired weighted average of previous local model updates~\cite{adikari2021CompressingTemporalCorrelation,chen2022SparsifiedGradientDifferences};
     \item linear combination of previous model updates with combination weights optimized by the least-squares criterion (\textit{i.e.}, minimizing the MSE distortion)~\cite{edin2024TemporalPredictiveCoding,azam2022RecyclingModelUpdates}.
\end{itemize}

Particularly, in~\cite{edin2024TemporalPredictiveCoding}, the prediction relies on a window of previously transmitted model updates $\set{\hat{\vect{g}}\ts{t-t'}}_{t'=1,\dots,h}$, which are combined through a linear predictor defined as
\begin{equation}
    \check{\vect{g}}_\textrm{lin}\ts{t}(\vect{a}) = \sum_{t'=1}^{h} a_{t'}\hat{\vect{g}}\ts{t-t'}.
\end{equation}
Here, $\vect{a}=[a_1,\ldots,a_h]$ contains the prediction coefficients and $h$ is the memory size. The optimal coefficients $\vect{a}^*{}\ts{t}\in\reals^h$ are found by minimizing the reconstruction error between the current model update $\vect{g}\ts{t}$ and the prediction $\check{\vect{g}}_\textrm{lin}\ts{t}(\vect{a})$, \textit{i.e.},
\begin{equation}
    \vect{a}^*{}\ts{t}= \argmin_{\vect{a}\in\reals^h}\norm{\check{\vect{g}}_\textrm{lin}\ts{t}(\vect{a}) -\vect{g}\ts{t}}^2.
\end{equation}
The predictor can also be written in a matrix form $\check{\vect{g}}_{k,\text{lin}}\ts{t}(\vect{a}) = M\vect{a}$, where
$M=(\vect{g}_{k}\ts{t-1},\dots,\vect{g}_{k}\ts{t-h})\in\reals^{d\times h}$.
When $M$ has full column rank, the \gls{mse}-optimal prediction coefficients can be obtained in closed-form as $\vect{a}^*{}\ts{t}= (M\transpose M)^{-1}M\transpose \vect{g}\ts{t}$.
This corresponds to projecting $\vect{g}\ts{t}$ onto the subspace spanned by the columns of $M$.

\section{Spatial Correlation}\label{sec:correlation:spatial}
We define spatial correlation as the similarity between local model updates $\vect{g}_k\ts{t}$ produced by different clients $k\in[K]$.
A primary source of spatial correlation is the training data available to each client. When clients’ local datasets are drawn from similar distributions, spatial correlation tends to be high.
In \gls{fl}, local datasets are very often statistically heterogeneous and unbalanced, which affects the degree of spatial correlation that can be observed during the training procedure~\cite{li2020FedProx,li2019convergence}. %

\subsection{How to Evaluate the Degree of Spatial Correlation?}\label{sec:spatial:measure}\label{sec:measure correlation:svd}
\smallnegvspace
The degree of spatial correlation between two clients can be quantified by the cosine similarity of their local model updates, \emph{i.e.}, $\css(\vect{g}_{k}\ts{t},\vect{g}_{l}\ts{t})$,
where $k$ and $l$ are two client indices. When the model update vectors from multiple clients are highly correlated, it is possible to identify a common lower-dimensional subspace that efficiently represents these updates.
Note that exploiting spatial correlation locally is difficult for individual clients unless each one has precise knowledge of the common subspace. Therefore, it requires the assistance of the \gls{ps}, which has knowledge of all local model updates.

Again, we use the linear regression example to illustrate the effects of spatial correlation. We consider two clients $k$ and $l$, each with $N_k$ and $N_l$ data samples. The local loss functions are given by $f_k(\vect{x})=\frac{1}{N_k} (\vect{x}\transpose R_k\transpose R_k\vect{x} - 2\vect{y}_k\transpose R_k \vect{x} + \vect{y}_k\transpose \vect{y}_k)$, and $f_l(\vect{x})=\frac{1}{N_l} (\vect{x}\transpose R_l\transpose R_l\vect{x} - 2\vect{y}_l\transpose R_l\vect{x} + \vect{y}_l\transpose \vect{y}_l)$, respectively. The cosine similarity between their gradient vectors is
\begin{align}\label{eq:toy:spatial}
    \css(&\nabla f_k(\vect{x}),\nabla f_l(\vect{x}))\nonumber\\ =&\frac{(R_k\transpose R_k\vect{x} - R_k\transpose \vect{y}_k)\transpose(R_l\transpose R_l\vect{x} - R_l\transpose \vect{y}_l)}   { N_k N_l \norm{\nabla f_k(\vect{x}\ts{t})} \norm{\nabla f_l(\vect{x}\ts{t})} }\nonumber\\
    =&\frac{\vect{x}\transpose A\vect{x} + \vect{b}\transpose\vect{x} + c}  {\norm{\nabla f_k(\vect{x}\ts{t})} \norm{\nabla f_l(\vect{x}\ts{t})} },
\end{align}
where $A = \frac{R_k\transpose R_k}{N_k} \frac{R_l\transpose R_l}{N_l}$, $\vect{b}\transpose = \frac{-1}{N_kN_l}(\vect{y}_k\transpose R_k R_l \transpose R_l + \vect{y}_l\transpose R_l R_k\transpose R_k)$, and $c=\frac{1}{N_kN_l}(\vect{y}_k\transpose R_k  R_l \transpose \vect{y}_l)$.
The gradient of the quadratic loss is invariant under unitary transformations of the column space. Specifically, if $R_l = PR_k$ and $\vect{y}_l=P\vect{y}_k$, for any orthonormal $P\in\reals^{N_k\times N_k}$, then $\nabla f_l(\vect{x}) = R_l\transpose R_l \vect{x} - R_l\transpose \vect{y}_l= R_k\transpose P\transpose P R_k \vect{x} - R_k\transpose P \transpose P \vect{y}_k = R_k\transpose R_k \vect{x} - R_k\transpose \vect{y}_k = \nabla f_k(\vect{x})$. Since the cosine similarity between any vector and itself is one, it follows that $\css(\nabla f_k(\vect{x}),\nabla f_l(\vect{x})) = 1$ in this case.

By examining the row space further, we can identify two cases where $\css(\nabla f_k(\vect{x}),\nabla f_l(\vect{x})) = 0$.
First, if the row spaces of $R_l$ and $R_k$ are orthogonal, \emph{i.e.}, $R_kR_l\transpose = \zerovec{N_k\times N_l}$, the cosine similarity is zero, which corresponds to the scenario of extreme data heterogeneity.
Otherwise, we obtain a level hypersurface of the quadratic function at which the cosine similarity is zero. Note that this set of points includes the minimizers of $f_k(\vect{x})$ and $f_l(\vect{x})$.
In the remaining parameter space, the cosine similarity is nonzero, indicating the existence of potential correlations that could be exploited for compression.

Quantifying spatial correlation between more than two clients is possible using a \gls{svd}-based measure. To facilitate this, we generalize the method in \cref{sec:exploiting correlations} by considering the \emph{observation matrix} $\Phi\in\reals^{p\times q}$. The structure of $\Phi$ depends on the type of correlation we aim to measure.
To measure spatial correlation in \gls{fl}, we use $\Phi = (\vect{g}_{1}\ts{t},\ldots,\vect{g}_{K}\ts{t})\in\reals^{d\times K}$, which contains the local gradient vectors from all clients, available at the \gls{ps}.
To measure structural correlation, we take $\Phi=G_k\ts{t}\in\reals^{m\times n}$, as in \cref{sec:exploiting correlations}.

Applying \gls{svd} to the measurement matrix gives $\Phi = U\Sigma V\transpose$, where $\Sigma\in\reals^{p\times q}$ contains the singular values $\sigma_i,\ i\in[\minX{p,q}]$, sorted as $\sigma_1 \geq .. \geq \sigma_{\minX{p,q}}$, on its diagonal elements, and $U\in\reals^{p\times p}$ and $V\in\reals^{q\times q}$ are orthonormal matrices.
\begin{definition}[Conserved Energy Ratio]\label{def:energy conservation ratio}
    Given an observation matrix $\Phi\in\reals^{p\times q}$, and $r\in [0,\minX{p,q}]$, the energy conserved by the first $r$ singular values is defined as
    \begin{align}\label{eq:energy conservation ratio}
        \alpha_\mathrm{cer} =
        \sum_{i=1}^{r}\sigma_i^2 \Big/\mspace{-5mu}\sum_{i=1}^{\minX{p,q}}\sigma_i^2 = \norm{\Sigma_{\rm 1:r,1:r}}_F^2/\norm{\Sigma}_F^2.
    \end{align}
\end{definition}
Using \cref{def:energy conservation ratio}, we define a measurement function $\MeasureCorrSVD{\Phi,\beta_\text{\rm svd}}$, where the inputs are:
\begin{itemize}
    \item an observation matrix $\Phi\in\reals^{p\times q}$, and
    \item a fraction $\beta_\text{\rm svd} \in [0,1]$,
\end{itemize}
and the output is the conserved energy ratio $\alpha_\mathrm{cer}\in[0,1]$ for $r = \ceil{\beta_\text{\rm svd}\cdot\minX{p,q}}$.
This function quantifies the proportion of energy captured by a low-rank approximation using a fraction $\beta_\text{svd}$ of the leading singular values.
As an example, for some energy threshold $\alpha_\text{svd}\in [0,1]$, we can consider sufficient correlation to be present if $\MeasureCorrSVD{\Phi,\beta_\text{svd}} = \alpha_\mathrm{cer} \geq \alpha_\textrm{svd}$.

While this heuristic rule can be applied, other metrics such as the effective rank~\cite{roy2007EffectiveRank} could provide a more principled way to quantify the extent of low-rank structure.

\subsection{Exploiting Spatial Correlation using Subspace Projection}\label{sec:spatial:exploiting}
\smallnegvspace
In addition to measuring the correlation degree, \gls{svd} can also be used to extract a subspace onto which we project the model update, thereby reducing its dimensionality.
After collecting local model updates from all clients, the \gls{ps} performs \gls{svd} on the measurement matrix $\Phi = (\vect{g}_{1}\ts{t},\ldots,\vect{g}_{K}\ts{t})\in\reals^{d\times K}$ as $\Phi=U\Sigma V\transpose$. Then, we define an operator $\NarrowSVD{\Phi,\alpha_\text{svd}}$, which, given inputs
\begin{itemize}
    \item an observation matrix $\Phi\in\reals^{p\times q}$, and
    \item a threshold $\alpha_\text{svd}\in[0,1]$,
\end{itemize}
returns the truncated SVD components
\begin{itemize}
    \item $\Sigma_r=\Sigma_{1:r,1:r}\in\reals^{r\times r}$,
    \item $U_r= U_{:,1:r}\in\reals^{p\times r}$, and
    \item $V_r=V_{:,1:r}\in\reals^{q\times r}$,
\end{itemize}
where
$r$ is selected as the smallest integer such that the conserved energy ratio (see \cref{def:energy conservation ratio}) is above the target value $\alpha_\text{\rm svd}$.
The \gls{ps} needs to broadcast $U_r$ (and possibly $V_r$) to all clients so that they can perform model update compression by subspace projection in subsequent iterations.
For example, when using  $U_r\in\reals^{d\times r}$ to compress the model update vector $\vect{g}_k\ts{t}\in\reals^d$, we have the compressed model update as
\begin{align}\label{eq:svd:general compression}
    \tilde{\vect{g}}_k\ts{t} = U_r\transpose\vect{g}_k\ts{t},
\end{align}
which reduces the number of transmitted elements from $d$ to $r$.
After the \gls{ps} receives $\tilde{\vect{g}}_k\ts{t}$, it can recover an approximate model update by
\begin{align}\label{eq:svd:general decompression}
    \hat{\vect{g}}_k\ts{t} = U_r\tilde{\vect{g}}_k\ts{t}= U_rU_r\transpose\vect{g}_k\ts{t}.
\end{align}

\begin{remark}
   The SVD-based low-rank approximation method presented in \cref{sec:exploiting correlations} projects the current model update matrix $G$ onto its top-$r$ singular subspace.
   The subspace projection method presented in this section uses a fixed subspace with previously learned basis from an earlier iteration. The difference is that exploiting spatial correlation among clients requires server-side processing, \textit{i.e.}, the \gls{ps} collects local updates to extract subspace information for client-side compression in later iterations.
\end{remark}

\smallnegvspace

\begin{table}[!tb]
\newcommand\tabyes{\checkmark}
\newcommand\tabno{---}
\newcommand{\rt}[1]{${}^{\text{#1}}$}
\caption{classification of recent compression schemes used in \gls{fl}.}\label{tab:classification}
\newcolumntype{C}[1]{>{\centering\arraybackslash}m{#1}}
\begin{tabular}{C{4.4cm}C{0.9cm}C{0.9cm}C{0.6cm}}
    & \textbf{\footnotesize Structural} & \textbf{\footnotesize Temporal} & \textbf{\footnotesize Spatial}\\
    \toprule
    ATOMO~\cite{wang2018ATOMO} & \tabyes & \tabno & \tabno \\\midrule
    LAG~\cite{chen2018LAG} &\tabno&\tabyes&\tabno\\\midrule
    Recycling~\cite{azam2022RecyclingModelUpdates}& \tabno & \tabyes & \tabno \\
    \midrule
    FedSketch ~\cite{rothchild2020FetchSGD} & \tabyes & \tabno & \tabno \\
    \midrule
    L-GreCo ~\cite{alimohammadi2023lgreco} & \tabyes & \tabno & \tabno \\
    \midrule
    \hspace*{-1mm}Temporal sparsification mask~\cite{ozfatura2021TemporalSparsification,sun2022TimeCorrelatedSparsificationOTA} & \tabyes & \tabyes\rt{1} & \tabyes \\
    \midrule
    Gradient difference~\cite{mishchenki2024CompressedGradientDifferences,richtarik2021EF21} & \tabyes & \tabyes\rt{1} & \tabno \\
    \midrule
    Momentum-style prediction~\cite{adikari2021CompressingTemporalCorrelation,chen2022SparsifiedGradientDifferences} & \tabno & \tabyes\rt{1} & \tabno \\
    \midrule
    Temporal predictive coding~\cite{edin2024TemporalPredictiveCoding} & \tabno & \tabyes & \tabno  \\
    \midrule
    SVDFed~\cite{wang2023SVDFed} & \tabno & \tabyes\rt{1} & \tabyes \\
    \midrule
    GradiVeQ~\cite{yu2018GradiVeQ} & \tabyes\rt{1} & \tabyes\rt{1} & \tabyes\rt{2} \\
    \midrule
    FedOComp~\cite{xue2022FedOComp} & \tabno & \tabyes\rt{1} & \tabyes\rt{2} \\
    \midrule
    PowerSGD~\cite{vogels2019PowerSGD,makkuva2024LASER} & \tabyes & \tabyes & \tabyes\rt{2} \\
    \midrule
    FedPara~\cite{hyeon2021fedpara} & \tabyes & \tabno &\tabno \\
    \midrule
    SVD-resuse~\cite{hao2024flora,zhao2024galore} & \tabyes & \tabyes\rt{1} & \tabno \\
    \midrule
    Proposed AdaSVDFed & \tabno & \tabyes & \tabyes \\
    \midrule
    Proposed PCAFed & \tabyes & \tabyes & \tabyes \\
    \bottomrule
\end{tabular}
\centering
\\[1mm]{\footnotesize \rt{1}Implicit usage.\rt{2}Considers the aggregate.%
}
\end{table}

\section{Benefits and Limitations of Subspace Projection Methods}\label{sec:improving by resuing}
While the idea of using subspace projection for model update compression in \gls{fl} is not new, our correlation-based taxonomy provides a unified lens for understanding the underlying design principles. In this section, we will classify existing studies into different categories, and discuss the benefits and limitations of these methods in terms of communication cost, computational, and memory overhead.

\subsection{Classification of Existing Compression Methods}
\cref{tab:classification} gathers and classifies some existing correlation-aware compression methods, depending on the sources of correlation they rely on. Next, we discuss some of them in more detail.

\subsubsection{
    Temporal Sparsification Mask
}
The works in~\cite{ozfatura2021TemporalSparsification,sun2022TimeCorrelatedSparsificationOTA} consider sparsification in the standard parameter basis and are therefore classified as exploiting structural correlation. These schemes leverage temporal correlation by retaining part of the sparsification mask across consecutive iterations, implicitly assuming temporal correlation. Moreover,~\cite{sun2022TimeCorrelatedSparsificationOTA} is also classified as exploiting spatial correlation because the portion of the mask preserved between iterations is shared across all clients.
\subsubsection{
    GradiVeQ
}
The work in~\cite{yu2018GradiVeQ} adopts an alternative transmission protocol, known as ring-all-reduce, which provides all clients with access to the aggregated model update once communication is complete. The update is divided into model update slices, similarly to the approach used in this paper. The method alternates between two phases: a training phase and a compression phase. During the training phase, each client stores model updates, after which a covariance matrix is computed. Temporal correlation is exploited by using \gls{pca}, and the basis matrix learned from the first slice is reused for all subsequent slices. We classify this use of structural and temporal correlations as implicit, since a single basis matrix is reused across slices and across multiple iterations.

\subsubsection{
    SVDFed
}\label{sec:methods:svdfed}
The work in~\cite{wang2023SVDFed} employs server-assisted compression to capture spatial correlation among clients using \gls{svd}. It uses
 $\Phi = (\vect{g}_{1}\ts{t},\ldots,\vect{g}_{K}\ts{t})$, and performs compression using a truncated left orthonormal matrix $U_r$.

These prior studies typically assume that the sources of correlation they rely on always exist. In practice, algorithm designers should verify the strength of these correlations and adapt the compression scheme accordingly, rather than applying one-size-fits-all compressors.

\subsection{Communication Cost Reduction}

The main advantage of subspace projection as a compression method is the reduced communication cost. Instead of transmitting  $d$ (entire parameter vector) or $m$ (one slice of the model update matrix) elements, only $r$ elements need to be transmitted.
However, the basis matrix is needed for both the clients and the \gls{ps} to perform compression and decompression. Transmitting the basis matrix incurs additional communication cost,
and the amount of extra communicated information depends on whether the clients or \gls{ps} compute the basis matrices.
When using client-side processing, the basis matrices are computed at each client and communicated to the \gls{ps}, thus the communication cost grows linearly with the number of clients $K$. When using server-side processing, the basis matrices are computed at the \gls{ps} and sent back to the clients, which can be done by broadcast.

The break-even point in terms of the total number of communicated elements can be easily calculated for each compression method.
Consider an example where we use server-side processing and want to compress one model update slice of the model update matrix, $\vect{g}_{k,i}\ts{t}$, using $\NarrowSVD{(\vect{g}_{1,i}\ts{t},\ldots,\vect{g}_{K,i}\ts{t}), \alpha}$, for some $\alpha$.
If we reuse $U_r$ for $j$ consecutive iterations, the total communication cost is $mr+rj$, for transmitting $U_r$ and $U_r\transpose\vect{g}_{k,i}\ts{t}$.
Uncompressed model updates for the same number of iterations require transmitting $jm$ elements.
When $mr+rj <jm \iff j> \frac{mr}{m-r}$, reusing the basis vectors over multiple iterations reduces the total communication cost.
Nevertheless, this calculation ignores that compression will introduce distortion into the model updates, potentially affecting training performance.

For example, suppose consecutive gradient iterates maintain a cosine similarity of at least $c$, so the angle between them is at most $\theta = \arccos(c)$.
After $j$ iterations, the angle between $\vect{g}\ts{t}$ and $\vect{g}\ts{t-j}$ can be as large as $j\theta$, giving $\cssX{\vect{g}\ts{t}}{\vect{g}\ts{t-j}} \geq \cos(j\theta)$.
To guarantee temporal correlation, we require $j \leq \frac{2\pi}{\theta}$. If the break-even number of iterations $\frac{mr}{m-r}$ exceeds this limit, reusing the basis matrix for that many iterations is not guaranteed to be useful.

\subsection{Implicit Dependence on Temporal Correlation}
A limitation of server-side computation of basis matrices is that after the \gls{ps} collects local model updates from all clients in iteration $t$, the extracted basis matrices can only be used for update compression in the subsequent iterations $t+1, t+2,\ldots$. Consequently, such compression methods rely on the implicit assumption of temporal correlation, meaning that the subspace information extracted from earlier iterations can remain useful later. With weak temporal correlation, the basis matrix will fail to capture the model evolution in subsequent iterations, thus resulting in high compression distortion.

\subsection{Computational and Memory Overhead}
Using \gls{svd} or \gls{pca} for measuring correlation and designing compression schemes requires additional computations.
For example, if we want to verify the degree of structural correlation before applying low-rank approximation to compress a model update matrix $G_k\ts{t}\in\reals^{m\times n}$, we only need to compute its singular values, and not the full \gls{svd}. If $m\geq n$, computating singular values requiring $\ordo{2mn^2 + 2n^3}$ FLOPS, compared to $\ordo{6mn^2 + 20n^3}$ FLOPS for the complete \gls{svd} \cite[Section 8.6]{golub2013matrixcomp}.
Computing \gls{pca}, especially for tall matrices with $m\gg n$, by explicitly forming the covariance matrix may be inefficient. Instead, one can exploit the close relationship between \gls{pca} and \gls{svd} to obtain the principal components and singular values of the covariance matrix directly via an \gls{svd}~\cite{gerbrands1981relationships}.
Moreover, since \gls{pca} requires mean subtraction—incurring an additional cost of $m(n+1)$ FLOPS—it typically entails higher computational overhead than \gls{svd}.

Memory overhead originates from the need to store previous model update information, such as in temporal-correlation-aware compression design via predictive coding, which consumes storage resources. When the number of clients grows, this memory overhead may become a significant concern.

\newcommand{\stUPDATE}{\textnormal{UPDATE}}
\newcommand{\stPCA}{\textnormal{PCA}}
\newcommand{\stLocalSVD}{\textnormal{LocalPCA}}

\newcommand{\stSVDFED}{\textnormal{SPATIAL}}
\newcommand{\stLS}{\textnormal{PRED}}

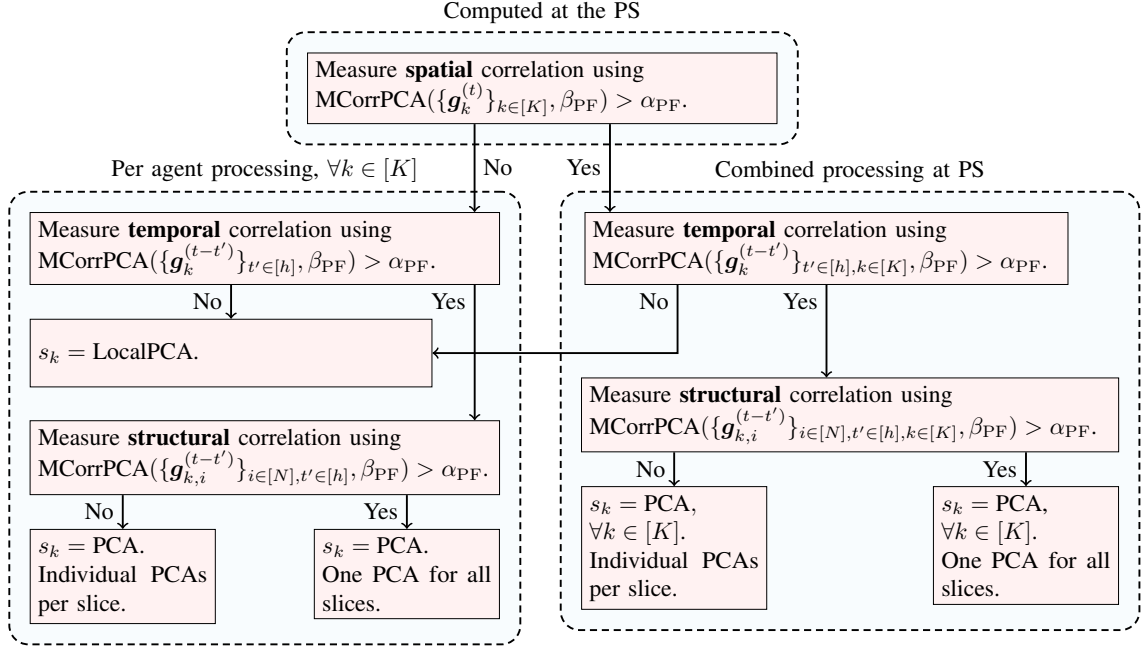
\begin{figure*}
    \centering
    \newcommand\bw{6.7cm}
\newlength{\blockvsep}
\setlength{\blockvsep}{0.5cm}
\scalebox{0.9}{
\begin{tikzpicture}
    \node[sysblock] (root) at (0,0) {\begin{minipage}{\bw} Measure \textbf{spatial} correlation using\newline $\MeasureCorrPCA{\set{\vect{g}_k\ts{t}}_{k\in[K]},\beta_\mathrm{PF}} > \alpha_\mathrm{PF}$. \end{minipage}};

    \node[sysblock, below left = 1.3cm and \dimexpr -\bw/2+0.5cm of root] (nosp) {\begin{minipage}{\bw}Measure \textbf{temporal} correlation using\newline $\MeasureCorrPCA{\set{\vect{g}_{k}\ts{t-t'}}_{t'\in[h]},\beta_\mathrm{PF}} > \alpha_\mathrm{PF}$. \end{minipage}};
    \coordinate (no start) at ($(root.south)+(-1cm,0)$);
    \draw[sysline] (no start) -- (nosp.north-|no start) node[pos=0.3,below right] {No};

    \node[sysblock, below left=\blockvsep and \dimexpr-\bw +0.75cm of nosp] (nosp note) {\begin{minipage}{\bw-1cm}$s_k = \stLocalSVD$.\end{minipage}};
    \draw[sysline] (nosp.south-|nosp note.north) -- (nosp note.north) node[pos=0.,below left]{No};

    \node[sysblock, below =1cm +2\blockvsep of nosp] (nosp yeste) {\begin{minipage}{\bw}Measure \textbf{structural} correlation using\newline $\MeasureCorrPCA{\set{\vect{g}_{k,i}\ts{t-t'}}_{i\in[N],t'\in[h]},\beta_\mathrm{PF}} > \alpha_\mathrm{PF}$. \end{minipage}};
    \draw[sysline] (nosp.south)+(\dimexpr \bw/2 -0.25cm,0) -- ($(nosp yeste.north)+(\dimexpr \bw/2 -0.25cm,0)$) node[pos=0,below left]{Yes};
    \node[sysblock, below left=\blockvsep and -2.75cm of nosp yeste] (nosp yeste nost) {\begin{minipage}{2.5cm}$s_k = \stPCA$. \\ Individual \glspl{pca} per slice. \end{minipage}};
    \draw[sysline] (nosp yeste.south-|nosp yeste nost.north) -- (nosp yeste nost.north) node[pos=0,below left]{No};
    \node[sysblock, below right =\blockvsep and -2.75cm of nosp yeste] (nosp yeste yesst) {\begin{minipage}{2.5cm} $s_k = \stPCA$. \\ One \gls{pca} for all slices. \end{minipage}};
    \draw[sysline] (nosp yeste.south-|nosp yeste yesst.north) -- (nosp yeste yesst.north) node[pos=0,below left]{Yes};

    \node[sysblock, below right= 1.3cm and \dimexpr -\bw/2+0.5cm of root] (yessp)  {\begin{minipage}{\dimexpr\bw+0.2cm}Measure \textbf{temporal} correlation using\newline $\MeasureCorrPCA{\set{\vect{g}_{k}\ts{t-t'}}_{t'\in[h],k\in[K]},\beta_\mathrm{PF}} > \alpha_\mathrm{PF}$.\end{minipage}};

    \coordinate (yes start) at ($(root.south)+(+1cm,0)$);
    \draw[sysline] (yes start) -- (yessp.north-|yes start) node[pos=0.3,below left] {Yes};

    \draw[sysline] (yessp.south)+ (-2.2cm,0) |- (nosp note.east) node[pos=0,below left]{No};%

    \node[sysblock, below right=1cm +2\blockvsep -1.5\baselineskip and \dimexpr -\bw-0.5cm of yessp] (yessp yeste) {\begin{minipage}{\bw+1cm}Measure \textbf{structural} correlation using\newline $\MeasureCorrPCA{\set{\vect{g}_{k,i}\ts{t-t'}}_{i\in[N],t'\in[h],k\in[K]},\beta_\mathrm{PF}} > \alpha_\mathrm{PF}$. \end{minipage}};
    \draw[sysline] (yessp.south) -- ($(yessp.south|-yessp yeste.north)$) node[pos=0,below left]{Yes};
    \node[sysblock, below left=\blockvsep and -2.75cm of yessp yeste] (yessp yeste nost) {\begin{minipage}{2.5cm}$s_k = \stPCA,$\\$\forall k\in[K]$.\\ Individual \glspl{pca} per slice.\end{minipage}};
    \draw[sysline] (yessp yeste.south-|yessp yeste nost.north) -- (yessp yeste nost.north) node[pos=0,below left]{No};
    \node[sysblock, below right =\blockvsep and -2.75cm of yessp yeste] (yessp yeste yesst) {\begin{minipage}{2.5cm} $s_k = \stPCA$,\\$\forall k\in[K]$.\\One \gls{pca} for all slices. \end{minipage}};
    \draw[sysline] (yessp yeste.south-|yessp yeste yesst.north) -- (yessp yeste yesst.north) node[pos=0,below left]{Yes};

    \begin{scope}[on background layer]
        \node[sysfitblock,fit=(nosp)(nosp note)(nosp yeste)(nosp yeste yesst)(nosp yeste nost),label=above:{Per agent processing, $\forall k\in[K]$},
        inner sep=3mm] (RNO) {};
        \node[sysfitblock,fit=(yessp)(yessp yeste)(yessp yeste nost)(yessp yeste yesst),label=above:{Combined processing at \gls{ps}},
        inner sep=3mm] (RYES) {};
        \node[sysfitblock,fit=(root),label=above:{Computed at the \gls{ps}},
        inner sep=3mm] (R) {};
        
    \end{scope}

\end{tikzpicture}
}
    \caption{The flowchat describing the state selection process in $\PCAFedUpdate{\set{\vect{g}_k\ts{t}}_{k=1,\dots,K},\alpha_\mathrm{PF},\beta_\mathrm{PF}}$.%
    }
    \label{fig:pcafed:flowchart}
\end{figure*}

\begin{algorithm}[!tbp]
\caption{AdaSVDFed.}\label{alg:AdaSVDFed}
\SetCommentSty{footnotesize}
\SetKwInOut{Input}{input}
\DontPrintSemicolon
\SetKwProg{MyBlock}{}{:}{}
\SetInd{-.0em}{0.3em}
\SetAlgoSkip{}
\DecMargin{2em}
Global state $s\in\set{\stUPDATE,\stSVDFED,\stLS}$\;
$M_k,M_{\mathrm{PS},k} \gets \zerovec{d\times h}, \forall k\in[K]$ \;
\For{$t=1,...,T$}
{
    \lIf{$t~\mathrm{mod}~T_u = 1$}{$s \gets \stUPDATE$.}

    Each client $k$ computes local model update vector $\vect{g}_k\ts{t}$\;

    \If{$s=\stUPDATE$}
    {
        Each client transmits $\vect{g}_k\ts{t}$ to the \gls{ps}\;
        \tcp{Procedure at \gls{ps}:}
        $G\ts{t} \gets (\vect{g}_1\ts{t}, ... , \vect{g}_K\ts{t})$\;
        \eIf{$\MeasureCorrSVD{G\ts{t}, \beta_\mathrm{SF} } > \alpha_\mathrm{SF}$}
        {
            \gls{ps} gets $U_r$ from $\NarrowSVD{G\ts{t}, \alpha_\mathrm{SF} }$ and broadcasts $U_r$ to all clients\;
            $s \gets \stSVDFED$\;
        }
        {
            Each client stores $\vect{m}_k \gets \vect{g}_k\ts{t}$\;
            \gls{ps} stores $\vect{m}_{{\rm PS},k}\gets \vect{g}_k\ts{t}$, $\forall k\in[K]$\;
            $s \gets \stLS$\;
        }
        $\hat{\vect{g}}_k\ts{t} \gets \vect{g}_k\ts{t}$\;
    }

    \ElseIf{$s=\stSVDFED$}
    {
        Each client transmits $\tilde{\vect{g}}_k\ts{t} \gets U_{r}\transpose\vect{g}_k\ts{t}$ to the \gls{ps}\;
        \gls{ps} reconstructs $\hat{\vect{g}}_k\ts{t} \gets U_{r} \tilde{\vect{g}}_k\ts{t}$\;
    }

    \ElseIf{$s=\stLS$}
    {

        \ForPar{\textnormal{clients} $k=1,...,K$}
        {
            \tcp{Client $k$ computes:}
            $\vect{a}_k\ts{t} \gets \argmin_\vect{a} \norm{\vect{g}_k\ts{t}-M_k\vect{a} }$\;
            $\vect{r}_k\ts{t} \gets \vect{g}_k\ts{t} - M_k\vect{a}_k\ts{t}$\;
            $\tilde{\vect{r}}_k\ts{t} \gets \cCX{\vect{r}_k\ts{t}}$\;
            Client $k$ transmits $\vect{a}_k\ts{t}$ and $\tilde{\vect{r}}_k\ts{t}$ to \gls{ps}\;
            \tcp{Both client $k$ and \gls{ps}:}
            Reconstruct $\hat{\vect{g}}_k\ts{t} \gets M_k\vect{a}_k\ts{t} + \textnormal{EXPAND}(\tilde{\vect{r}}_k\ts{t})$\;
            Update memory: $M_k \gets \bigl[\hat{\vect{g}}_k\ts{t},\, [\vect{M}_k]_\mathrm{:,:h-1}\bigr]$, $M_{\mathrm{PS},k} \gets \bigl[\hat{\vect{g}}_k\ts{t},\, [\vect{M}_k]_\mathrm{:,:h-1}\bigr]$\;

        }
    }

    \gls{ps} updates $ \vect{x}\ts{t+1} \gets \vect{x}\ts{t} + \sum_{k=1}^Kw_k\hat{\vect{g}}_k\ts{t} $ and transmits $\vect{x}\ts{t+1}$ to all clients\;
}
\end{algorithm}

\begin{algorithm}[!tbp]
\caption{PCAFed.}\label{alg:pcafed}
\SetKwInOut{Input}{input}
\DontPrintSemicolon
\SetKwProg{MyBlock}{}{:}{}
\SetInd{-.0em}{0.3em}
\SetCommentSty{footnotesize}
\SetAlgoSkip{}
\DecMargin{2em}
Client states $s_k\in\set{\stUPDATE, \stPCA,\stLocalSVD}$\;
\For{$t=1,...$}
{
    \lIf{$t~\mathrm{mod}~T_u = 1$}{Set $s_k \gets \stUPDATE$, $\forall k\in[K]$.}

    Each client $k$ computes local model update vector $\vect{g}_k\ts{t}$\;

    \tcp{State: \stUPDATE}
    \If{$s_k = \stUPDATE$ \textnormal{for all} $k\in[K]$}
    {
        Each client transmits $\vect{g}_k\ts{t}$ to the \gls{ps}\;\tcp{Procedure at \gls{ps}:}
        \gls{ps} updates states $s_k$ using $\PCAFedUpdate{\set{\vect{g}_k\ts{t}}_{k=1,\dots,K},\alpha_\mathrm{PF},\beta_\mathrm{PF}}$\;
        \For{\textnormal{clients} $k$ \textnormal{where} $s_k = \stPCA$}
        {
            \gls{ps} transmits $\vect{\mu}_{k,i}\ts{t}$, $Q_{k,i}\ts{t}$ to client $k$, $\forall i\in[n]$\;
        }
        $\hat{\vect{g}}_k\ts{t}\gets \vect{g}_k\ts{t}$, $\forall k\in[K]$\;
    }
    \Else
    {
        \tcp{State: \stPCA}
        \ForPar{\textnormal{clients} $k$ \textnormal{where} $s_k=\stPCA$}
        {
            \ForPar{\textnormal{slice} $i=1,...,n$}
            {
                Client $k$ transmits $\tilde{\vect{g}}_{k,i}\ts{t} \gets [Q_{k,i}\ts{t}]\transpose(\vect{g}_{k,i}\ts{t}-\vect{\mu}_{k,i}\ts{t})$ to \gls{ps}\;
                \gls{ps} reconstructs $\hat{\vect{g}}_{k,i}\ts{t} \gets Q_{k,i}\ts{t} \tilde{\vect{g}}_{k,i}\ts{t}$, $\forall i\in[n]$\;
            }
        }

        \tcp{State: \stLocalSVD}
        \ForPar{\textnormal{clients} $k$ \textnormal{where} $s_k=\stLocalSVD$}
        {
            \eIf{$\MeasureCorrPCA{\set{\vect{g}_{k,i}\ts{t}}_{i\in[n]},\beta_\mathrm{PF}} > \alpha_\mathrm{PF}$}
            {
                Client $k$ computes $Q_{k,r}\ts{t}$ using $\NarrowPCA{\set{\vect{g}_{k,i}\ts{t}}_{i\in[n]},\alpha_\mathrm{PF}}$\;
                Client $k$ transmits $Q_{k,r}\ts{t}$, $\vect{\mu}_{k}\ts{t}$, and $\tilde{\vect{g}}_{k,i}\ts{t} \gets [Q_{k,r}\ts{t}]\transpose(\vect{g}_{k,i}\ts{t}-\vect{\mu}_{k}\ts{t})$, $\forall i\in[n]$\;
                \gls{ps} reconstructs $\hat{\vect{g}}_{k,i}\ts{t} \gets Q_{k,r}\ts{t}\tilde{\vect{g}}_{k,i}\ts{t} + \vect{\mu}_{k}\ts{t}$, $\forall i\in[n]$\;
            }
            {
                \tcp{Insufficient exploitable correlation}
                Client $k$ transmits $\vect{g}_k\ts{t}$ to \gls{ps}\;%
                $\hat{\vect{g}}_k\ts{t} \gets \vect{g}_k\ts{t}$\;
            }
        }
    }

    \gls{ps} updates $ \vect{x}\ts{t+1} \gets \vect{x}\ts{t} + \sum_{k=1}^Kw_k\hat{\vect{g}}_k\ts{t} $ and transmits $\vect{x}\ts{t+1}$ to all clients\;
}
\end{algorithm}

\section{Adaptive Compression Designs with Multiple Sources of Correlation}
The three types of correlation may coexist in an \gls{fl} system. Thus, an efficient compression design can benefit from jointly considering multiple sources of correlation.
In this section, we introduce two adaptive and context-aware compression designs: \textit{AdaSVDFed} in \cref{sec:AdaSVDFed}, an augmented version of SVDFed, and \textit{PCAFed} in \cref{sec:PCAFed}, which can exploit all three types of correlation when present and switch between different compression modes depending on the degree of measured correlation.

\subsection{AdaSVDFed}\label{sec:AdaSVDFed}
We introduce AdaSVDFed in \cref{alg:AdaSVDFed}, which switches between \gls{svd}-based and predictive-coding-based compression modes depending on the measured spatial correlation.
Every $T_u$ iterations, the algorithm enters an update state $s=\stUPDATE$, where the \gls{ps} measures spatial correlation to decide whether SVDFed compression in \cite{wang2023SVDFed} or predictive-coding compression in \cite{edin2024TemporalPredictiveCoding} will be used for the next $T_u-1$ iterations.
Specifically, the switching decision is governed by $\beta_\mathrm{SF} \in[0,1]$ and $\alpha_\mathrm{SF} \in[0,1]$:
we set $s = \stSVDFED$ and use SVDFed compression (described in \cref{sec:methods:svdfed}) if no more than $\ceil{\beta_\mathrm{SF}\cdot\minX{K,d}}$ basis vectors are needed to capture at least a fraction $\alpha_\mathrm{SF}$ of the energy in $(\vect{g}_1\ts{t},\dots,\vect{g}_K\ts{t})$.
Otherwise, we set $s=\stLS$ to use the predictive‑coding mode (described in \cref{sec:correlation:temporal:exploit}), in which each client further compresses the prediction residual using $\cCX{}:\reals^d\rightarrow\reals^p$ (\emph{e.g.}, sparsification) and the \gls{ps} reconstructs the vector via the reverse operation $\textnormal{EXPAND}(\cdot):\reals^p\rightarrow\reals^d$.
The key advantage of AdaSVDFed is that it avoids transmitting the basis matrix $U_r$ from the \gls{ps} to the clients when the spatial correlation is low, thereby reducing downlink communication cost when the basis provides little compression benefit.

Note that both SVD and predictive coding methods already exist in the literature. With this AdaSVDFed algorithm, our goal is to demonstrate that introducing an adaptive switching mechanism can improve existing compression designs, as shown in \cref{sec:numerical} through simulation results.

\vspace{-0.5cm}
\subsection{PCAFed}\label{sec:PCAFed}
Until now, \gls{svd} has been the primary method considered for measuring and exploiting correlations.
An alternative option is to use \gls{pca}, as discussed in \cref{sec:evaluate correlations,sec:exploiting correlations}.
Let $\cG$ be a set of $p$-dimensional vectors, for instance, $\cG= \set{\vect{g}_{k,i}\ts{t-t'}}_{i\in[n],t'\in[h]}$ with $p = m$, which contains all model update slices (with lenght $m$) of client $k$ in multiple iterations. In this way, we can jointly measure the temporal and structural correlations.
Compute the mean $\vect{\mu} = 1/\abs{\cG}\sum_{\vect{g}'\in\cG}\vect{g}'\in\reals^{p}$ and the covariance
\begin{equation}
    C = \frac{1}{\abs{\cG}-1}\sum_{\vect{g}'\in\cG} (\vect{g}'-\vect{\mu})(\vect{g}'-\vect{\mu})\transpose\in\reals^{p\times p}.
\end{equation}
We find the principal values $\lambda_i,~i=1,\dots,p$, from the eigenvalue decomposition $C = Q\Lambda Q\transpose$, where $\Lambda =\diag(\lambda_1,\dots,\lambda_p)$ is the diagonal matrix of eigenvalues, ordered such that $\lambda_1\geq\dots\geq\lambda_p\geq0$, and $Q\in\reals^{p\times p}$ is an orthonormal matrix.
To measure the degree of correlation using PCA, we consider the approximate energy ratio defined as
\begin{align}\label{eq:pca:energy rule}
    \alpha_\mathrm{aer} = \frac{\norm{\vect{\mu}}_2^2 + \sum_{i=1}^{r}\lambda_i}{\norm{\vect{\mu}}_2^2 + \sum_{i=1}^{p}\lambda_i}. %
\end{align}
Then, using \eqref{eq:pca:energy rule}, we define a measurement function $\MeasureCorrPCA{\cG,\beta_\textrm{pca}}$, where the inputs are:
\begin{itemize}
    \item a set of $p$-dimensional vectors $\cG$, and
    \item a fraction $\beta_\textrm{pca} \in [0,1]$,
\end{itemize}
and the output is the approximate energy ratio $\alpha_\mathrm{aer} \in [0,1]$ for $r = \ceil{\beta_\textrm{pca}\cdot p}$.
This function provides an approximate method for quantifying the energy captured by a low-rank subspace.
As an example, for a threshold $\alpha_\textrm{pca}\in [0,1]$, we consider correlation to be sufficient if $\MeasureCorrPCA{\cG,\beta_\textrm{pca}} = \alpha_\mathrm{aer} \geq \alpha_\textrm{pca}$.

Using \eqref{eq:pca:energy rule} also allows us to design a compression block $\NarrowPCA{\cG,\alpha_\textrm{pca}}$, which, given inputs
\begin{itemize}
    \item a set of $p$-dimensional vectors $\cG$, and
    \item a threshold $\alpha_\textrm{pca}\in(0,1]$,
\end{itemize}
returns
\begin{itemize}
    \item the truncated $Q_r = Q_{:,1:r}\in\reals^{p\times r}$, and
    \item the mean $\vect{\mu}\in\reals^{p}$,
\end{itemize}
where $r$ is selected as the smallest integer such that $\MeasureCorrPCA{\cG,\beta_\textrm{pca}} \geq \alpha_\textrm{pca}$.
Note that $r=0$ is possible if the mean value contains enough energy. Using the mean vector $\vect{\mu}$ and $Q_r$, a client can compress a model update slice $\vect{g}_{k,i}\ts{t}$ by applying \eqref{eq:pca:compression}, and the server decompresses it using \eqref{eq:pca:decompression}.

\begin{figure*}[!tbh]
    \centering
    \begin{subfigure}[t]{0.95\linewidth}
        \begin{subfigure}[t]{0.23\linewidth}
            \centering
            \includegraphics[width=\linewidth]{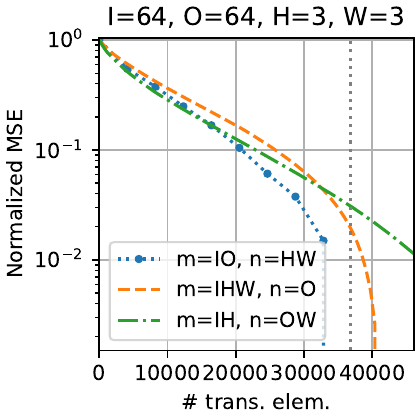}
            \caption{layer1.1.conv2 (5th)}\label{fig:structural:reset:112}
        \end{subfigure}
        \hfill
        \begin{subfigure}[t]{0.23\linewidth}
            \centering
            \includegraphics[width=\linewidth]{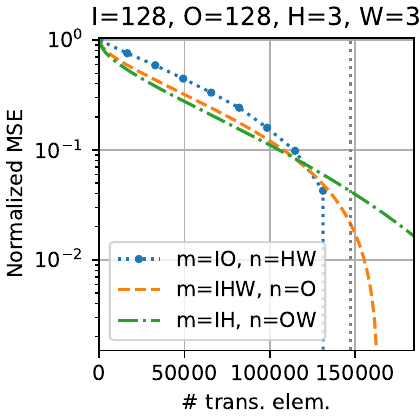}
            \caption{layer2.1.conv2 (9th)}\label{fig:structural:reset:212}
        \end{subfigure}
        \hfill
        \begin{subfigure}[t]{0.23\linewidth}
            \centering
            \includegraphics[width=\linewidth]{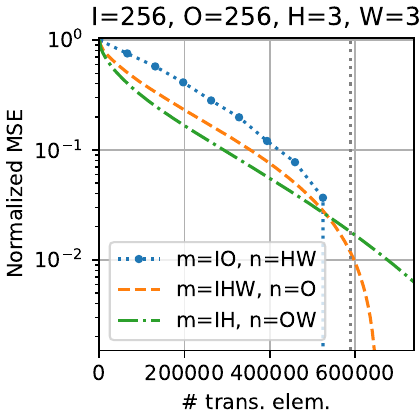}
            \caption{layer3.1.conv2 (13th)}\label{fig:structural:reset:312}
        \end{subfigure}
        \hfill
        \begin{subfigure}[t]{0.23\linewidth}
            \centering
            \includegraphics[width=\linewidth]{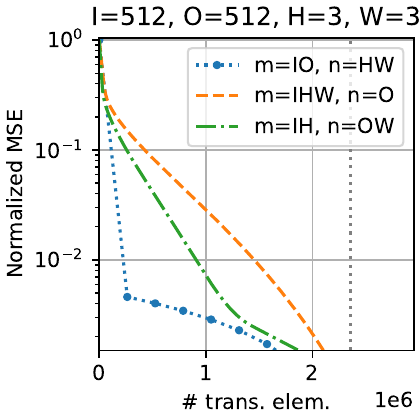}
            \caption{layer4.1.conv2 (17th)}\label{fig:structural:reset:412}
        \end{subfigure}
    \end{subfigure}
    \caption{Normalized \gls{mse} of the model update matrix approximation $\hat{G}\ts{100} \in \reals^{m \times n}$ for different ResNet18 layers (total depth: 18), and their depth in the network, as a function of the number of transmitted elements over the client-to-\gls{ps} link.
    When $\min(m, n)$ is small, the number of possible choices is limited, and we explicitly mark the attainable points in these cases.
    The vertical gray dotted line marks $mn=IOHW$ transmitted elements (number of elements without compression). $\tau = 32, \gamma =0.001$.}
    \label{fig:structural:several_layers}
        \label{fig:structural:resenet:112 212 312 412}
\end{figure*}

\begin{figure}[!tbp]
    \centering
    \includegraphics[scale=.66]{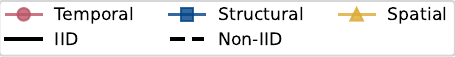}\\
    \begin{subfigure}[t]{0.49\linewidth}
        \centering
        \includegraphics[width=\linewidth]{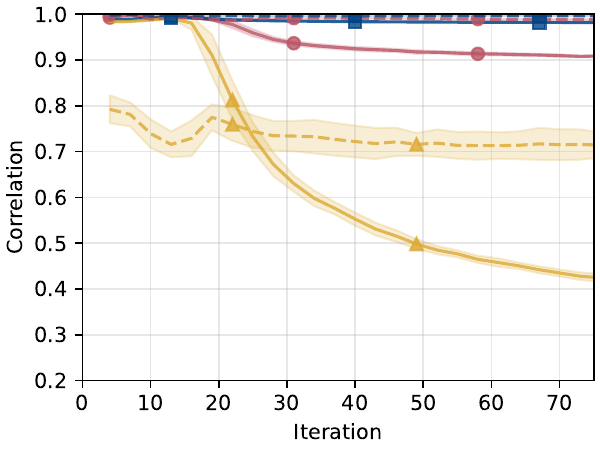}
        \caption{LeNet.}
        \label{fig:correlation over time:lenet}
    \end{subfigure}
    \begin{subfigure}[t]{0.49\linewidth}
        \centering
        \includegraphics[width=\linewidth]{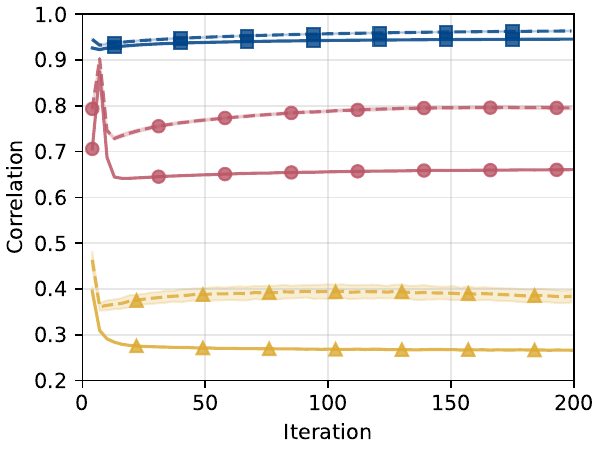}
        \caption{ResNet.}
        \label{fig:correlation over time:resnet}
    \end{subfigure}
    \caption{Measured correlation during training across all layers, where each layer's influence is proportional to its number of parameters. Under the non-IID setting, each client holds at most 4 of 10 labels. We measure structural correlation by $\MeasureCorrPCA{\set{\vect{g}_{k,i}\ts{t}}_{i\in[n]},0.2}$, temporal correlation by $\MeasureCorrPCA{\set{\vect{g}_{k}\ts{t-t'}}_{t'=0,\dots,4}, 0.2}$, which is averaged over the $K=20$ clients. Spatial correlations are measured as $\MeasureCorrPCA{\set{\vect{g}_{k}\ts{t}}_{k\in[K]},0.2}$. The results are averaged over 10 runs, and the fill shows $\pm$ one standard deviation. $\tau=32$, $\gamma=0.001$.}
    \label{fig:correlation over time}
    \vspace*{1mm}
\end{figure}

We introduce PCAFed in \cref{alg:pcafed}, an adaptive compression design that actively measures and exploits the three types of correlation.
Each client $k$ has a state $s_k$, $k=1,\ldots,K$, that determines the compression mode it currently uses.
Every $T_u$ iterations, the algorithm sets all clients to the update state $s_k=\stUPDATE$, $k=1,\dots, K$, to measure the degree of correlation and apply the most suitable compression mode for the subsequent $T_u-1$ iterations.

Let $\PCAFedUpdate{\set{\vect{g}_k\ts{t}}_{k=1,\dots, K},\alpha_\mathrm{PF},\beta_\mathrm{PF}}$ be the mechanism that the \gls{ps} uses to update the compression mode for all clients, where $\alpha_\mathrm{PF}\in[0,1]$ and $\beta_\mathrm{PF}\in[0,1]$. To do so, the \gls{ps} measures correlation according to the flowchart in \cref{fig:pcafed:flowchart} to decide the next state of each client.
In the absence of temporal correlation, the update mechanism sets $s_k = \stLocalSVD$, which requires no additional computations at the \gls{ps}. A client in this state applies \gls{pca}-based compression to its local model update each iteration if sufficient structural correlation is measured within the update.
In the presence of temporal correlation, the update method can exploit structural and/or spatial correlation, which correspond to the four cases that use $s_k = \stPCA$ in \cref{fig:pcafed:flowchart}. In all four cases, the \gls{ps} computes the mean $\vect{\mu}_{k,i}\ts{t}$ and the basis matrix $Q_{k,i}\ts{t}$, but each case uses different information depending on the present correlation(s):
\begin{itemize}
    \item \textbf{Spatial and structural correlation:} \\Collect $n$ model update slices from $K$ clients in $h$ iterations.  Compute \textbf{one} \gls{pca} to obtain
    $\vect{\mu}\ts{t},Q\ts{t} = \NarrowPCA{\set{\vect{g}_{k,i}\ts{t-t'}}_{i\in[n],t'\in[h],k\in[K]},\alpha_\mathrm{PF}}$, and assign
    $\vect{\mu}_{k,i}\ts{t}= \vect{\mu}\ts{t}, Q_{k,i}\ts{t} = Q\ts{t},\forall i \in [n], \forall k\in[K]$.
    \item \textbf{Spatial correlation:}\\
    Collect one model update slice from $K$ clients in $h$ iterations.
    Compute $n$ \glspl{pca} (one for each slice) to obtain
    $\vect{\mu}_{i}\ts{t},Q_{i}\ts{t}=\NarrowPCA{\set{\vect{g}_{k,i}\ts{t-t'}}_{t'\in[h],k\in[K]},\alpha_\mathrm{PF}},\forall i \in [n]$,
    and assign
    $\vect{\mu}_{k,i}\ts{t}= \vect{\mu}_{i}\ts{t}, Q_{k,i}\ts{t} = Q_{i}\ts{t},\forall i \in [n], \forall k\in[K]$.
    \item \textbf{Structural correlation:}\\
    Collect $n$ model update slices from one client in $h$ iterations.
    Compute $K$ \glspl{pca} (one for each client)  to obtain
    $\vect{\mu}_{k}\ts{t},Q_{k}\ts{t}=\NarrowPCA{\set{\vect{g}_{k,i}\ts{t-t'}}_{i\in[n],t'\in[h]},\alpha_\mathrm{PF}}$, $\forall k\in[K]$,
    and assign
    $\vect{\mu}_{k,i}\ts{t}= \vect{\mu}_{k}\ts{t}, Q_{k,i}\ts{t} = Q_{k}\ts{t},\forall i \in [n], \forall k\in[K]$.
    \item \textbf{Only temporal correlation:}\\
    Collect one model update slice from one client in $h$ iterations.
    Compute $nK$ \glspl{pca} to obtain
    $\vect{\mu}_{k,i}\ts{t}, \ Q_{k,i}\ts{t} = \NarrowPCA{\set{\vect{g}_{k,i}\ts{t-t'}}_{t'\in[h]},\alpha_\mathrm{PF}},\forall i \in [n],\allowbreak \forall\nobreak{}k\nobreak{}\in\nobreak{}[K]$.
\end{itemize}
The \gls{ps} then transmits the corresponding mean vector(s) and basis matrix (matrices) to each client $k$ with $s_k = \stPCA$.

\begin{remark}
    One drawback of this algorithm is the high computational complexity. While impractical in any real implementation, it serves as an example to illustrate how correlation-awareness can be incorporated into the compression design.
\end{remark}
\begin{figure*}[!tbh]
    \centering
    \begin{minipage}[c]{0.95\linewidth}
    \newcommand{\temporcorrscle}{0.90}
    \begin{subfigure}[t]{0.49\linewidth}
        \centering
        \includegraphics[width=\temporcorrscle\linewidth]{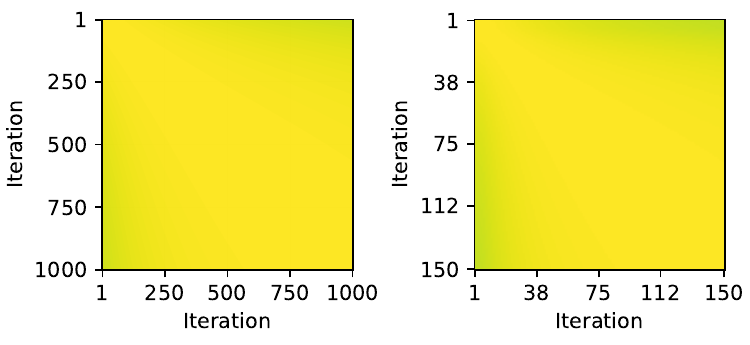}
        \caption{ Linear model. \\Left: $\tau=1 , \gamma=0.01$. Right: $\tau=32 , \gamma=0.0025$}
        \label{fig:res:temporal:x}
    \end{subfigure}
    \begin{subfigure}[t]{0.49\linewidth}
        \centering
        \includegraphics[width=\temporcorrscle\linewidth]{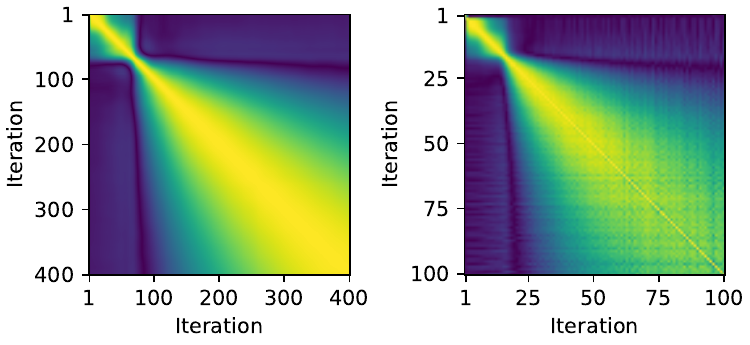}
        \caption{ LeNet, layer: fc1.\\ Left: $\tau=1 , \gamma=0.05$. Right: $\tau=32 , \gamma=0.001$}
        \label{fig:res:temporal:fc1}
    \end{subfigure}
    \begin{subfigure}[t]{0.49\linewidth}
        \centering
        \includegraphics[width=\temporcorrscle\linewidth]{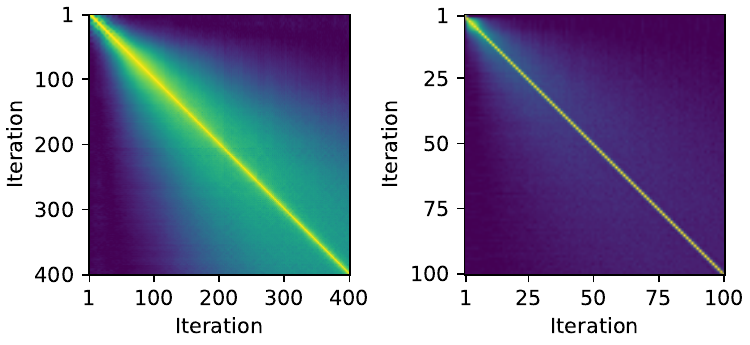}
        \caption{ResNet18, layer: layer4.0.conv1. \\ Left: $\tau=1 , \gamma=0.01$. Right: $\tau=32 , \gamma=0.001$}
        \label{fig:res:temporal:layer4.0.conv1}
    \end{subfigure}
    \begin{subfigure}[t]{0.49\linewidth}
        \centering
        \includegraphics[width=\temporcorrscle\linewidth]{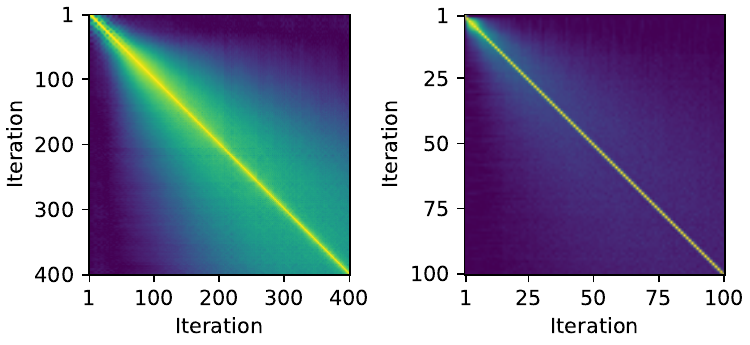}
        \caption{ResNet18, layer: layer4.0.conv2. \\ Left: $\tau=1 , \gamma=0.01$. Right: $\tau=32 , \gamma=0.001$}
        \label{fig:res:temporal:layer4.0.conv2}
    \end{subfigure}
    \end{minipage}
    \begin{minipage}[c]{0.042\linewidth}
        \includegraphics[width=\linewidth]{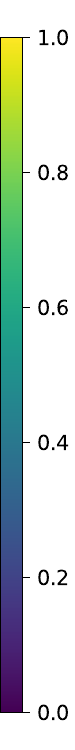}\vspace*{1cm}
    \end{minipage}
    \caption{Temporal correlation between model updates over time evaluated using $\abs{\css(\vect{g}\ts{t_1},\vect{g}\ts{t_2})}$.}%
    \label{fig:res:temporal}
\end{figure*}

\section{Numerical Experiments}\label{sec:numerical}
\smallnegvspace
In this section, we present simulation results to quantify the degrees of structural, temporal, and spatial correlations, using the metrics discussed in the previous sections.\footnote{The code will be publicly available after the review process.}
We consider three learning tasks with different learning models:
\begin{itemize}
    \item A binary classification task, using a linear model fitted via logistic regression. Each data sample $(\vect{\xi}_i,y_i)\in \reals^d\times\{-1,1\}$ is a feature-label pair, and the sample-wise loss $\ell(\vect{x};(\vect{\xi}_i,y_i)) = \log(1+e^{-y_i\vect{\xi}_i\transpose\vect{x}})$. %
    We use the w8a dataset from~\cite{chang2011libsvm}, which contains $\num{49749}$ samples with $d=300$ features.
    \item An image classification task on the MNIST dataset~\cite{lecun2010mnist}, which has $\num{50000}$ training samples, using a small convolutional neural network LeNet~\cite{lecun1998Lenet}, with $\num{44426}$ trainable parameters.
    \item An image classification task on the CIFAR10 dataset~\cite{krizhevsky2016tinyimages}, which has $\num{50000}$ training samples, using a larger convolutional neural network ResNet18~\cite{he2016DeepResidualLearning}, with $\num{11181642}$ trainable parameters.
\end{itemize}

The training data samples are evenly distributed among the clients in the system. We examine both IID settings, in which each client possesses an equal share of all class labels, and non‑IID settings, where clients hold
a subset of available classes.
Each client applies mini-batch \gls{sgd} and adapts the batch size $\abs{\cD_k\ts{t,i}}$ based on the number of local \gls{sgd} steps $\tau$ such that $\abs{\cD_k\ts{t,i}}\cdot\tau=|\mathcal{D}_k|$.
For the image classification tasks, we apply heavy-ball momentum with parameter $0.9$ in the local \gls{sgd} steps, and a weight decay of factor $10^{-4}$.

For learning models whose parameters naturally divide into layers, \emph{e.g.}, the layers of a neural network, we apply compression independently to each such layer.
To identify where the largest gains can be achieved by introducing redundancy-based compression, we focus on layers that account for a significant portion of the total parameters.
For the LeNet model trained on the MNIST dataset, we consider the fully connected layer {fc1}, with shape $[256, 120]$, as it contains approximately 70\% of the model's parameters. For the ResNet18 model trained on the CIFAR-10 dataset, we focus on the convolutional layers {layer4.0.conv1}, with shape $[256, 512, 3, 3]$, and {layer4.0.conv2}, with shape $[512, 512, 3, 3]$, which account for approximately 11\% and 21\% of the total parameters, respectively.

\begin{figure*}[!tbp]
    \centering
    \begin{subfigure}[t]{0.49\linewidth}
        \centering
        \includegraphics[scale=.6]{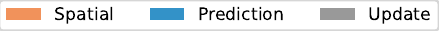}
        \\
        \includegraphics[width=0.48\linewidth]{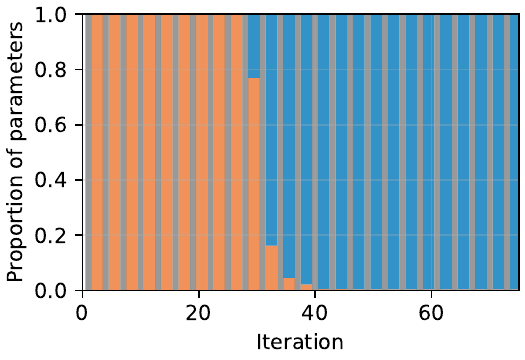}
        \hfill
        \includegraphics[width=0.48\linewidth]{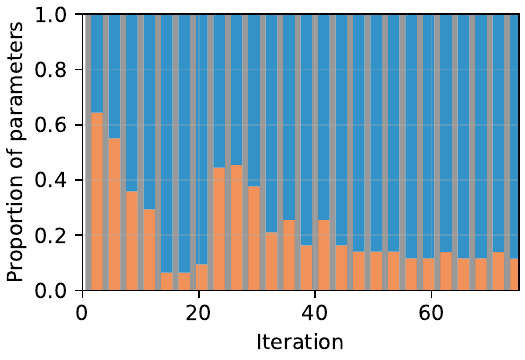}
        \caption{LeNet, using AdaSVDFed. IID to the left and non-IID to the right.}
        \label{fig:stateswitching:adasvdfed:lenet}
    \end{subfigure}
    \hfill
    \begin{subfigure}[t]{0.49\linewidth}
        \centering
        \hspace*{-2mm}\includegraphics[scale=.56]{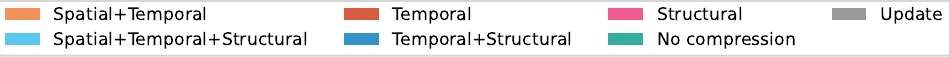}
        \\
        \includegraphics[width=0.48\linewidth]{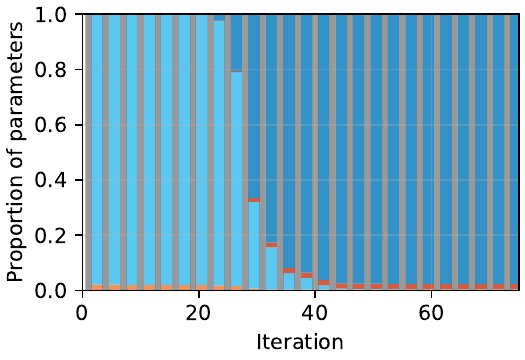}
        \hfill
        \includegraphics[width=0.48\linewidth]{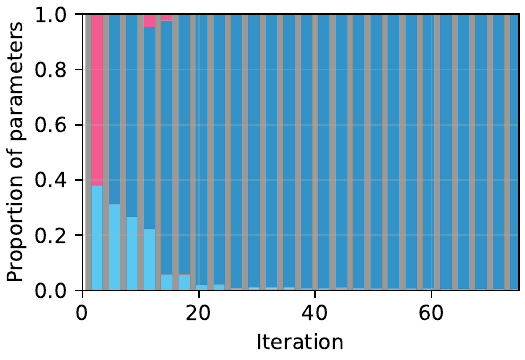}
        \caption{LeNet, using PCAFed. IID to the left and non-IID to the right.}
        \label{fig:stateswitching:pcafed:lenet}
    \end{subfigure}
    \\[5pt]
    \begin{subfigure}[t]{0.49\linewidth}
        \centering
        \includegraphics[scale=.6]{legend_adasvdfed.pdf}
        \\
        \includegraphics[width=0.48\linewidth]{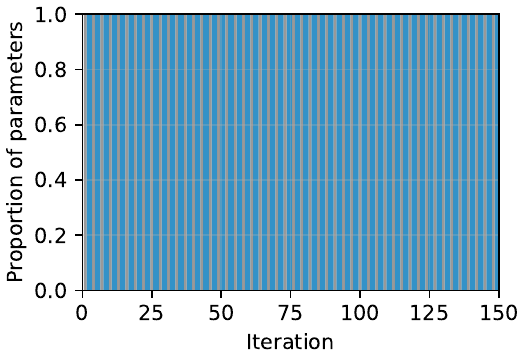}
        \hfill
        \includegraphics[width=0.48\linewidth]{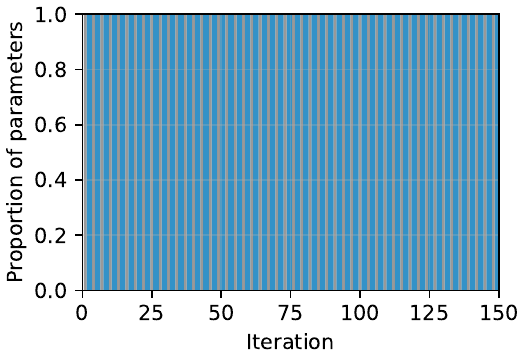}
        \caption{Resnet18, using AdaSVDFed. IID to the left and non-IID to the right.}
        \label{fig:stateswitching:adasvdfed:resnet}
    \end{subfigure}
    \hfill
    \begin{subfigure}[t]{0.49\linewidth}
        \centering
        \hspace*{-2mm}\includegraphics[scale=.56]{legend_pcafed.pdf}
        \\
        \includegraphics[width=0.48\linewidth]{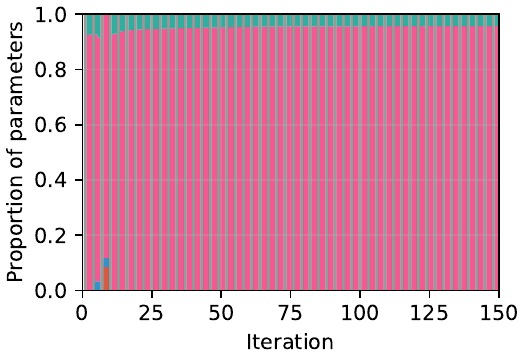}
        \hfill
        \includegraphics[width=0.48\linewidth]{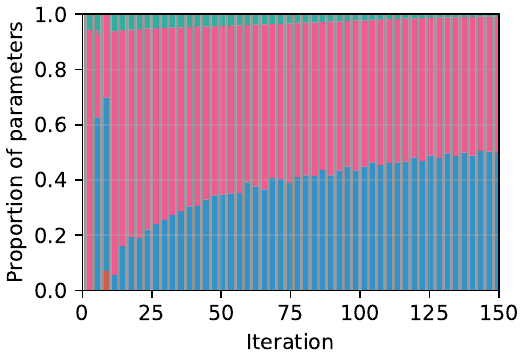}
        \caption{Resnet18, using PCAFed. IID to the left and non-IID to the right. }
        \label{fig:stateswitching:pcafed:resnet}
    \end{subfigure}

    \caption{Proportion of parameters operating under each state during training, averaged across $K=20$ clients. Under the non-IID setting, each client holds at most 4 of 10 labels. The ``No compression'' reported for PCAFed is when the system is in \stLocalSVD{}, but insufficient structural correlations are found, so the raw update is transmitted. $\alpha=0.8$, $\beta = 0.2$, $\tau=32$, $\gamma=0.001$. The results are average over 10 runs. }
    \label{fig:stateswitching}
\end{figure*}

\subsection{Correlation Evolution}

\Cref{fig:correlation over time} shows how the correlation strength evolves during training with $K=20$ clients. The results indicate that the correlation varies over time and depends on both the model architecture and the data distribution. In \cref{fig:correlation over time:lenet}, we see that after an initial phase with high correlation, the IID scenario experiences a substantial decline in spatial correlation and a minor drop in temporal correlation during the later part of training. While initially weaker, the non-IID setting maintains a more stable spatial correlation.
Meanwhile, \cref{fig:correlation over time:resnet} shows that the structural correlation is strong regardless of data distribution, while the degree of temporal and spatial correlation is larger in the non-IID setting.

\subsection{Impact of Parameter Reshaping on Structural Correlation}\label{sec:structure:reshape}

The structural correlation measures in \cref{fig:correlation over time} are for one specific mapping from convolutional kernel to model update matrix, that is, $n = OH$ and $m = IW$.

In \cref{fig:structural:several_layers}, we explore different mappings---choosing $m$ and $n$ as various combinations of $O$, $I$, $H$, and $W$---and compare how the \gls{mse} of the low-rank approximation $\hat{G}\ts{t}$ reconstructed at the \gls{ps} varies with the number of transmitted elements.
We present four convolutional layers from different depths in the ResNet18 model: layer1.1.conv2 (5th layer), layer2.1.conv2 (9th layer), layer3.1.conv2 (13th layer), and layer4.1.conv2 (17th layer), in \cref{fig:structural:reset:112,fig:structural:reset:212,fig:structural:reset:312,fig:structural:reset:412}, respectively.
We find that no single mapping is optimal across all layers, assigning $I$ and $O$ to different axes generally yields a lower \gls{mse} for the same number of transmitted elements.

\subsection{Impact of Multiple Local Steps on Temporal Correlation }\label{sec:res:temporal}
We use cosine similarity to measure the strength of temporal correlation at a single client when using different numbers of local \gls{sgd} steps across three learning tasks.
For the linear model in (\cref{fig:res:temporal:x}), a high degree of temporal correlation is observed for both small ($\tau = 1$) and large ($\tau = 32$) numbers of local steps.
From \cref{fig:res:temporal:fc1}, we observe strong temporal correlation during two distinct phases of the LeNet training process. However, for ResNet18 (\cref{fig:res:temporal:layer4.0.conv1,fig:res:temporal:layer4.0.conv2}), the correlation diminishes as the number of local steps increases.
These results suggest that the model update trajectory can be more complex for certain models and datasets, resulting in weaker temporal correlation.

\subsection{State Evolution}
In \cref{fig:stateswitching}, we illustrate how the system switches between different states and compression modes when implementing \cref{alg:AdaSVDFed,alg:pcafed}.
When training LeNet under an IID data distribution, we find that in the early iterations, both AdaSVDFed and PCAFed tend to select compression modes that exploit multiple sources of correlation (e.g., spatial+temporal or spatial+temporal+structural).
In the non-IID case, both algorithms use compression modes less frequently that rely less on spatial correlation.
However, for ResNet18, the state selection is hindered by the weaker temporal correlation seen in \cref{fig:correlation over time:resnet,fig:res:temporal}, especially for the IID case, which limits how often both algorithms can exploit multiple correlations simultaneously.
For AdaSVDFed in \cref{fig:stateswitching:adasvdfed:resnet}, the results reveal heavy use of the predictive coding mode, where a prediction and a compressed residual are transmitted.
For PCAFed in \cref{fig:stateswitching:pcafed:resnet}, the results indicate that $\stLocalSVD$ is used most frequently, where raw model updates are transmitted due to a lack of correlation. In contrast, the slightly stronger temporal correlation in the non-IID scenario is sufficient to enable joint exploitation of spatial and temporal correlations. In the early iterations in \cref{fig:correlation over time:resnet}, we see a spike in temporal correlation, which is believed to cause the temporary use of temporal correlation in \cref{fig:stateswitching:pcafed:resnet}.

\begin{figure}[!tbp]
    \centering
    \includegraphics[scale=.66]{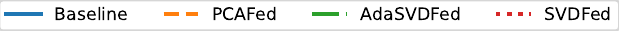}\\
    \begin{subfigure}[t]{0.49\linewidth}
        \centering
        \includegraphics[width=\linewidth]{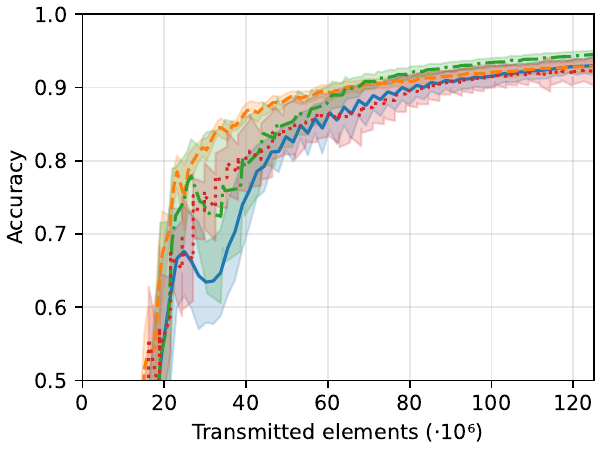}
        \caption{LeNet.}
        \label{fig:accuracy vs transmitted:lenet}
    \end{subfigure}
    \begin{subfigure}[t]{0.49\linewidth}
        \centering
        \includegraphics[width=\linewidth]{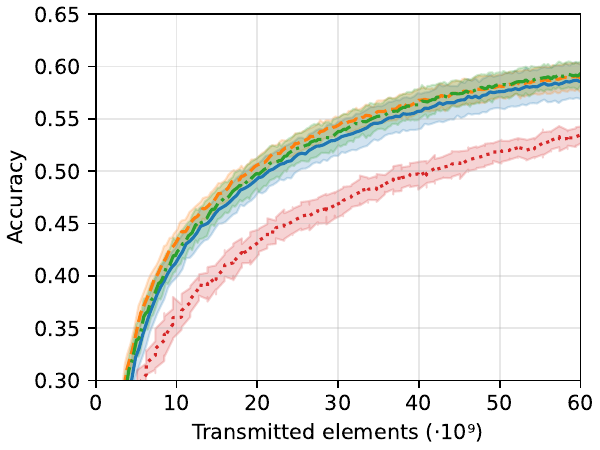}
        \caption{ResNet.}
        \label{fig:accuracy vs transmitted:resnet}
    \end{subfigure}
    \caption{Training performance versus total number of transmitted elements for non-IID data distribution. Under the non-IID setting, each client holds at most 4 of 10 labels. The results are average over 10 runs, and the fill shows $\pm$ one standard deviation. $\tau=32$, $\gamma=0.001$.}
    \label{fig:accuracy vs transmitted}
\end{figure}

\begin{table*}[ht]
    \caption{Training performance of \cref{alg:pcafed,alg:AdaSVDFed} compared to SVDFed with fixed update interval $T_u$, and a baseline without compression. Under the non-IID setting, each client holds at most 4 of 10 labels.  The reported values are averaged over 10 runs. $\tau=32,\ \gamma=0.001,\ K=20$. }
    \label{tab:performance}
    \centering
    \begin{tabular}{cccccSSS}

\textbf{Model} & \textbf{\shortstack{Dataset, Distr.}} & \textbf{Compression Method} & \textbf{Update Freq.} & \textbf{\# Iter.} &\textbf{\shortstack{\# Trans. Elem.\\(Agents $\rightarrow$ \gls{ps})}} &\textbf{\shortstack{\# Trans. Elem.\\(\gls{ps} $\rightarrow$ Agents)}} &\textbf{\shortstack{\# Total \\Trans. Elem.}} \\ 

\toprule

\multirow{8}*{LeNet} 

\multirow{8}*{LeNet} 
& \multirow{4}*{\shortstack{\footnotesize {\small MNIST, IID}\\[1mm]Until reaching 95\%\\validation accuracy}} 
  & None & --- & 50 & 44.07e6 & 44.07e6 & 88.14e6 \\ %
& & PCAFed & $T_S=3$ & 52 & 17.74e6 & 32.91e6 & 50.65e6 \\ %
& & AdaSVDFed & $T_S=3$ & 53 & 16.67e6 & 42.39e6 & 59.05e6 \\ %
& & SVDFed & $T_S=3$ & 67 & 19.78e6 & 39.55e6 & 59.33e6 \\ %
\cmidrule{2-8}
& \multirow{4}*{\shortstack{\footnotesize {\small MNIST, non-IID}\\[1mm]Until reaching 90\%\\validation accuracy}}
  & None & --- & 43 & 38.30e6 & 38.30e6 & 76.59e6 \\ %
& & PCAFed & $T_S=3$ & 47 & 15.08e6 & 35.33e6 & 50.42e6 \\ %
& & AdaSVDFed & $T_S=3$ & 43 & 13.95e6 & 48.98e6 & 62.93e6 \\ %
& & SVDFed & $T_S=3$ & 86 & 25.89e6 & 53.04e6 & 78.93e6 \\ %

\midrule
\multirow{8}*{ResNet18} 
& \multirow{4}*{\shortstack{\footnotesize {\small CIFAR10, IID}\\[1mm]Until reaching 65\%\\validation accuracy}}
  & None & --- & 87 & 19.41e9 & 19.41e9 & 38.81e9 \\ %
& & PCAFed & $T_S=3$ & 105 & 10.74e9 & 23.52e9 & 34.26e9 \\ %
& & AdaSVDFed & $T_S=3$ & 122 & 10.07e9 & 27.19e9 & 37.27e9 \\ %
& & SVDFed & $T_S=3$ & 185 & 13.90e9 & 67.52e9 & 81.42e9 \\ %
\cmidrule{2-8}

& \multirow{4}*{\shortstack{\footnotesize {\small CIFAR10, non-IID}\\[1mm]Until reaching 55\%\\validation accuracy}}
  & None & --- & 81 & 18.20e9 & 18.20e9 & 36.39e9 \\ %
& & PCAFed & $T_S=3$ & 104 & 9.85e9 & 23.28e9 & 33.13e9 \\ %
& & AdaSVDFed & $T_S=3$ & 111 & 9.16e9 & 24.80e9 & 33.97e9 \\ %
& & SVDFed & $T_S=3$ & 202 & 15.11e9 & 54.72e9 & 69.83e9 \\ %

\bottomrule
\end{tabular}

\end{table*}

\subsection{End-to-End Training Performance of Context-Aware Adaptive Compression Design}

At last, we compare the training performance of \cref{alg:pcafed,alg:AdaSVDFed} to a baseline without compression and SVDFed.\footnote{we consider a simplified version of SVDFed with periodic updating of the subspace-defining matrix.} We apply per-layer compression, and fix the update frequency $T_u = 3$.
For AdaSVDFed, we use $h=5$, $\alpha_\mathrm{SF}=0.8$, $\beta_\mathrm{SF} = 0.2$, and let $\cCX{}$ be a $\text{Top-}5\%$ compression operator, \textit{i.e.}, preserving the $5\%$ elements with largest magnitude (rounded up to the nearest integer).
For PCAFed, we use $h=5$, $\alpha_\mathrm{PF}=0.8$ and $\beta_\mathrm{PF}=0.2$.
We note that the parameters have not been fine-tuned and may not be optimal for performance. Finding optimal parameters for the compression schemes is outside the scope of this paper and is left for future work.

Generally, we see in \cref{tab:performance} that all compression schemes reduce the number of elements in the client-to-\gls{ps} links, at the cost of transmitting more elements in the \gls{ps}-to-client direction. In some cases, the \gls{ps}-to-client transmission
accounts for the majority of the total communication cost.

The PCAFed scheme shows strong performance on the tasks at hand, and its savings relative to other schemes can be attributed to its ability to adapt to changing correlations during training. In the scenario with stronger correlations (\textit{i.e.}, when training LeNet), we observe comparable iteration complexity. For the larger model (\textit{i.e.}, ResNet18), the iteration complexity increases slightly, but the increase is less pronounced than for the other schemes.

Comparing the number of transmitted elements between SVDFed and AdaSVDFed demonstrates the potential benefit of adaptive designs: in all cases where spatial correlation is weak, we observe a significant reduction in the number of communicated elements in the \gls{ps}-to-client link. We also observe a significant increase in iteration complexity for SVDFed when temporal correlation is weak, \textit{e.g.}, when training ResNet18.

\section{Conclusion and Future Work}\label{sec:conclusion}
\smallnegvspace
We categorize three types of correlation: structural, temporal, and spatial, which are inherent in \gls{fl} algorithms. Using unified terminology and notations, we also describe several correlation-exploiting schemes for compressing local model updates in the literature and clarify their underlying design principles.
We present numerical examples demonstrating that significant correlations can emerge in certain scenarios and can be leveraged to develop more efficient compression designs. These correlations vary across learning tasks and among different components (\textit{e.g.}, layers of a neural network) of the learned model.
Therefore, it is necessary to measure the degree of correlation (either locally or centrally) and adapt compression strategies to the correlation structures of the model components.

We design two adaptive compression algorithms that dynamically switch between different correlation‑exploiting modes: one that augments an existing design with a dynamic switching mechanism, and another that can jointly exploit structural, temporal, and spatial correlations when present. Through numerical experiments, we demonstrate that these adaptive algorithms outperform their non‑adaptive counterparts in terms of communication cost.
We leave the search for optimal hyperparameters and the further development of optimized, scalable, and correlation‑centric designs for future work.

Several other aspects remain to be investigated further, such as considering compression optimized for metrics other than \gls{mse}, exploring connections between commonly used bounds and our correlation measurement, and examining the effects of additional algorithmic add‑ons (\emph{e.g.}, error compensation).

\printbibliography
\end{document}